\documentclass{aa} 
\usepackage{amsmath}
\usepackage{graphicx}
\usepackage{url}
\usepackage{twoopt}
\usepackage{natbib}
\bibpunct{(}{)}{;}{a}{}{,} 
\usepackage{amssymb}
\usepackage{float}
\usepackage{color}
\usepackage{epstopdf}
\usepackage{ctable}
\usepackage{multirow}
\usepackage{stmaryrd}
\usepackage[draft]{hyperref}
\hypersetup{colorlinks=true,linkcolor=blue,citecolor=blue,filecolor=blue,urlcolor=blue}

\newcommand{\asec}[1]{{#1}$^{\prime\prime}$}

\begin{document}

\title{Extremely deep 150 MHz source counts from the LoTSS Deep Fields}
\titlerunning{LoTSS Deep Field Source Count}

\author{
S.~Mandal\thanks{E-mail: mandal@strw.leidenuniv.nl}\inst{1}\and 
I.~Prandoni\thanks{E-mail: prandoni@ira.inaf.it}\inst{2}\and
M.~J.~Hardcastle\inst{3}\and
T.~W.~Shimwell\inst{1,4}\and
H.~T.~Intema\inst{1,5}\and
C.~Tasse\inst{6,7}\and  
R.~J.~van~Weeren\inst{1}\and
H.~Algera\inst{1}\and
K.~L.~Emig\inst{1}\and
H.~J.~A.~R\"ottgering\inst{1}\and 
D.~J.~Schwarz\inst{8}\and
T.~M.~Siewert\inst{8}\and
P.~N.~Best\inst{9}\and
M.~Bonato\inst{2,10,11}\and
M.~Bondi\inst{2}\and
M.~J.~Jarvis\inst{12,13}\and
R.~Kondapally\inst{9}\and
S.~K.~Leslie\inst{1}\and\\
V.~H.~Mahatma\inst{14}\and
J.~Sabater\inst{9,15}\and
E.~Retana-Montenegro\inst{16}\and
W.~L.~Williams\inst{1}
}
\institute{Leiden Observatory, Leiden University, PO Box 9513, NL-2300 RA Leiden, The Netherlands \and
INAF - IRA, via P.~Gobetti 101, I-40129 Bologna, Italy \and
ASTRON, the Netherlands Institute for Radio Astronomy, Postbus 2, NL-7990 AA Dwingeloo, The Netherlands \and
Centre for Astrophysics Research, School of Physics, Astronomy and Mathematics, University of Hertfordshire, College Lane, Hatfield AL10 9AB, UK \and
International Centre for Radio Astronomy Research -- Curtin University, GPO Box U1987, Perth, WA 6845, Australia \and
GEPI \& USN, Observatoire de Paris, Université PSL, CNRS, 5 Place Jules Janssen, 92190 Meudon, France \and 
Department of Physics \& Electronics, Rhodes University, PO Box 94, Grahamstown, 6140, South Africa\and
Fakult\"at f\"ur Physik, Universit\"at Bielefeld, Postfach 100131, 33501 Bielefeld, Germany\and
SUPA, Institute for Astronomy, Royal Observatory, Blackford Hill, Edinburgh, EH9 3HJ, UK\and
Italian ALMA Regional Centre, Via Gobetti 101, I-40129, Bologna, Italy\and
INAF-Osservatorio Astronomico di Padova, Vicolo dell'Osservatorio 5, I-35122, Padova, Italy.\and
Astrophysics, Department of Physics, Keble Road, Oxford, OX1 3RH, UK 
Department of Physics \& Astronomy\and
University of the Western Cape, Private Bag X17, Bellville, Cape Town, 7535, South Africa\and
Th\"uringer Landessternwarte, Sternwarte 5, 07778 Tautenburg, Germany\and
STFC UK Astronomy Technology Centre, Royal Observatory, Blackford Hill, Edinburgh, EH9 3HJ, UK\and
Astrophysics \& Cosmology Research Unit, School of Mathematics, Statistics \& Computer Science, University of KwaZulu-Natal, Durban, 3690, South Africa
}
\authorrunning{S.~Mandal et al.}
\date{\today}
\date{Accepted XXX. Received YYY; in original form ZZZ}
\label{firstpage}
\abstract{With the advent of new generation low-frequency telescopes, such as the LOw Frequency ARray (LOFAR), and improved calibration techniques, we have now started to unveil the sub GHz radio sky with unprecedented depth and sensitivity. The LOFAR Two Meter Sky Survey (LoTSS) is an ongoing project in which the whole northern radio sky will be observed at 150 MHz with a sensitivity better than 100 $\mu$Jy beam$^{-1}$ at a resolution of \asec{6}. Additionally, deeper observations are planned to cover smaller areas with higher sensitivity. The Lockman Hole, the Bo\"otes and the Elais-N1 regions are among the most well known northern extra-galactic fields, and the deepest of the LoTSS Deep Fields so far. We exploit these deep observations to derive the deepest radio source counts at 150~MHz to date. Our counts are in broad agreement with those from the literature, and show the well known upturn at $\leq$ few mJy, mainly associated with the emergence of the star-forming galaxy population. More interestingly, our counts show for the first time a very pronounced drop around S$\sim$2 mJy, which results in a prominent `bump' at sub-mJy flux densities. Such a feature was not observed in previous counts' determinations (neither at 150 MHz nor at higher frequency). While sample variance can play a role in explaining the observed discrepancies, we believe this is mostly the result of a careful analysis aimed at deblending confused sources and removing spurious sources and artifacts from the radio catalogues. This `drop and bump' feature cannot be reproduced by any of the existing state-of-the-art evolutionary models, and appears to be associated with a deficiency of AGN at intermediate redshift ($1<z<2$) and an excess of low-redshift ($z<1$) galaxies and/or AGN.}
\maketitle

\section{Introduction}
Large-area radio surveys are very important for statistical studies of radio source populations, addressing astrophysical properties and cosmological evolution  of radio galaxies, quasars and starburst galaxies. In the past, several wide-area radio surveys were carried out at low radio frequencies, such as the Cambridge Surveys (3C, 4C, 6C, and 7C at around 160 MHz: \citealt{edge59}, \citealt{bennett62}, \citealt{pilkington65}, \citealt{gower67}, \citealt{baldwin85}). However, calibration of low-frequency radio data is challenging due to the direction-dependent, time-varying effects of the ionosphere that affects both the amplitude and the phase of the radio signal. Since these effects are only prominent in the MHz regime, 
the focus of  wide-area/all-sky radio surveys switched to around 1 GHz  in the last decades, resulting in the NRAO VLA Sky Survey (NVSS:\citealt{condon98}), the Sydney University Molonglo Sky Survey (SUMSS: \citealt{sumss03}) and the Faint Images of the Radio Sky at Twenty-Centimeters (FIRST) survey (\citealt{becker95}; \citealt{white97}). 
 
The higher sensitivity and higher spatial resolution of surveys at GHz frequencies also allowed us to probe deeper and deeper flux densities, and today we have several deep surveys covering degree-scale fields, and sensitive to the sub-mJy and $\mu$Jy radio populations (see e.g. \citealt{prandoni00a,prandoni00b,prandoni06}; \citealt{hopkins03}; \citealt{schinnerer04,schinnerer07}; \citealt{hales14a}; \citealt{smolcic17}; \citealt{prandoni18}). 
After many years of studies, it is now well established that the sub-mJy radio population has a composite nature. Radio-loud (RL) active galactic nuclei (AGN) are dominant down to 1.4 GHz flux densities of 200-300 $\mu$Jy  and star-forming galaxies (SFGs) become dominant below about 100-200 $\mu$Jy (\citealt{smolcic08}; \citealt{bonzini13}; \citealt{prandoni18}; \citealt{bonato20}). A significant fraction of the sources below 100 $\mu$Jy can also show signatures of AGN activity in the host galaxy at other bands (IR, optical, X-ray), but rarely display the large-scale radio jets and lobes typical of classical radio galaxies. Most of them are unresolved or barely resolved on a few arcsec scale, i.e. on scales similar to the host galaxy size. 
The origin of the radio emission in these (so-called radio-quiet) AGN is debated: it may come from star formation in the host galaxy (\citealt{padovani11,padovani15}, \citealt{bonzini13,bonzini15}; \citealt{ocran17}; \citealt{bonato17}) or from low-level nuclear activity (\citealt{white15, white17}; \citealt{maini16}; \citealt{herrera16,herrera17}; \citealt{hartley19}). 
Most likely, such AGN are composite systems where star formation and AGN-triggered radio emission  co-exist over a wide range of relative contributions (e.g. \citealt{delvecchio17}). This  scenario is also supported by the modeling work of \citet[see also Macfarlane et al. in prep.]{mancuso17}. 

Being sensitive to SFGs up to the epoch of the peak of their activity ($z \sim 2-3$), and reaching for the first time the dominant radio-quiet (RQ) AGN population, deep radio surveys probing the $\mu$Jy regime can be used as a very important dust/gas-obscuration-free tool to study both AGN activity and star formation and how they evolve with cosmic time. However, to overcome uncertainties introduced by low statistics, cosmic variance effects \citep{heywood13} and other systematics \citep{condon12}, deep-radio surveys that cover wide areas ($>>$ 1 deg$^2$) and have multi-band ancillary data are needed. Such wide--area surveys are also useful to investigate the role of environment in driving the growth of galaxies and SMBH, and to better trace rare radio source populations.

With the advent of a new generation of low-frequency telescopes and better data processing techniques we can now revisit the radio sky at low-frequency. With the Murchison Widefield Array (MWA; \citealt{lonsdale09}), \cite{wayth15} have carried out the GaLactic and Extragalactic All-sky MWA survey (GLEAM; \citealt{hurley-walker17}), reaching a sensitivity of a few mJy beam$^{-1}$ at a resolution of a few arcminutes. The GMRT has significantly improved the low-frequency view of the radio sky in terms of sensitivity and angular resolution. This has already been shown in a few low-frequency surveys centred around 150~MHz (e.g.: \citealt{ishwarachandra07}, \citealt{sirothia09}, \citealt{intema11}, \citealt{intema17}). 

The Low Frequency Array (LOFAR; \citealt{haarlem13}) is one of the key pathfinders to the Square Kilometre Array (SKA). Most of the LOFAR antennas are based in the Netherlands, with baseline lengths ranging from 100 meters to 120 km. Additional remote stations are located throughout various countries in Europe. The longest baseline of LOFAR can provide a resolution of \asec{0.3} at 150 MHz. The combination of LOFAR's large field of view, wide range of baseline lengths, and large fractional bandwidth makes it a powerful instrument for performing large area and deep sky surveys. The LOFAR Two Meter Sky Survey (LoTSS) is an an ongoing project in which the whole northern sky is observed with a sensitivity better than 100$\mu$Jy beam$^{-1}$ at the resolution of \asec{6} allowed by the Dutch LOFAR stations. The first data release (DR1) is described by \cite{shimwell17} and \cite{shimwell19}. The LoTSS also includes deeper observations of a number of pre-selected regions, where the aim is to eventually reach an rms depth of 10 $\mu$Jy beam$^{-1}$ at 150 MHz \citep{rottgering11}.  In order to scientifically exploit these more sensitive surveys (collectively known as LoTSS Deep Fields), complementary multi-wavelength data are necessary, most notably to identify the host galaxies of the extra-galactic radio sources and determine their redshift. For this reason observations were focused on fields with the highest quality multi-wavelength data available. The Lockman Hole, the Bo\"otes and the European Large-Area ISO Survey-North 1 (ELAIS-N1) fields are the deepest of the LoTSS Deep Fields so far (see \citealt{tasse20}; \citealt{sabater20}; respectively paper I and II of this series). 
All have rich multi-wavelength ancillary data, covering a broad range of the electromagnetic spectrum, from X-ray to radio bands. 

\ctable[topcap,center,star,
caption = {An overview of the statistical properties of the three LoTSS Deep Fields.},
label = tab:image-properties,
doinside=\scriptsize,
]{c | c c c c | c c c | c c c}
{

}
{
\FL Field & R.A. & DEC. & Obs. Time & $\sigma_{\rm c}$ &  Area$^{\rm full}$ & N$_{\rm S}^{\rm raw}$ & $\sigma_{\rm med}^{\rm full}$ &  Area$^{\rm masked}$ & N$_{\rm S}^{\rm final}$ & $\sigma_{\rm med}^{\rm masked}$ 
\NN  & (hh:mm:ss) & (dd:mm:ss) & (hr) & ($\mu$Jy beam$^{-1}$) &  (deg$^2$) &  & ($\mu$Jy beam$^{-1}$) &  (deg$^2$) &  & ($\mu$Jy beam$^{-1}$)
\ML LH  & 10:47:00.0 & +58:04:59.0   & 112   & 22 & 25.0 & 50112 & 42 & 10.3 & 31162 & 31
\NN Boo & 14:32:00.0 & +34:30:00.0 & 80 & 32 & 26.5 & 36767 & 60 & 8.6 & 19179  & 44
\NN EN1 & 16:11:00.0 & +55:00:00.0 & 164 & 17 & 24.3 & 69954 & 33 & 6.7 & 31610 & 23
\LL
\multicolumn{11}{l}{{\bf Notes:} the columns are as follows:  pointing centre (R.A. and DEC.); total observing time; RMS noise reached at the center of the image ($\sigma_c$); }\\
\multicolumn{11}{l}{area covered by  the raw catalogue (Area$^{\rm full}$), number of sources in the raw catalogue (N$^{\rm raw}_{\rm S}$) and median RMS noise in the area covered by the }\\
\multicolumn{11}{l}{raw catalogue ($\sigma^{\rm full}_{\rm med}$); same parameters for the final catalogue (Area$^{\rm masked}$, N$^{\rm final}_{\rm S}$, $\sigma^{\rm masked}_{\rm med}$).}

}


The Lockman Hole (LH hereafter) is one of the best studied extragalactic regions of the sky. It is characterized by a very low column density of Galactic HI (\citealt{lockman86}) making it an ideal field to study extragalactic sources with deep observations in the mid-IR/FIR/sub-mm (\citealt{lonsdale03}; \citealt{mauduit12}, \citealt{oliver12}),  optical/NIR (\citealt{muzzin09}; \citealt{foto12}; \citealt{hildebrandt16}), and X-ray (\citealt{polletta06}, \citealt{brunner08}). A variety of radio surveys cover limited areas within the LH region, at several frequencies. The widest deep radio survey so far consists of a 6.6 deg$^2$. 1.4 GHz mosaic obtained with the Westerbork (WSRT) telescope (1$\sigma$ sensitivity $\sim 10$ $\mu$Jy beam$^{-1}$; \citealt{prandoni18}). We refer to \cite{prandoni18} for a comprehensive summary of the available multi-frequency and multi-band coverage in this region (see also \citealt{kondapally20}, paper III of this series).

The Bo\"otes (Boo hereafter) field was originally targeted as  part of the NOAO Deep Wide Field Surveys (NDWFS; \citealt{jannuzi99}) which covers $\sim$ 9 deg$^2$ in the optical and near infrared (K) bands. Ancillary data is available for this field including X-ray (\citealt{murray05}; \citealt{kenter05}), UV (GALEX; Martin et al. 2003), and mid-infrared (\citealt{eisenhardt04}). Radio observations have also been carried out at 153 MHz with the GMRT (\citealt{intema11}; \citealt{williams13}), at 325 MHz with the VLA (\citealt{croft08}, \citealt{coppejans15}) and at 1.4 GHz with the WSRT (\citealt{devries02}).

The Elais-N1 (EN1 hereafter) field has deep multi-wavelength (0.15$\mathrm{\mu}$m - 250$\mathrm{\mu}$m) data taken as part of many different surveys (optical: the Panoramic Survey Telescope and Rapid Response System; Pan-STARRS \citep{ps1} and Hyper-Suprime-Cam Subaru Strategic Program (HSC-SSP) survey, u-band: Spitzer Adaptation of the Red-sequence Cluster Survey; SpARCS: \citealt{muzzin09}, UV: Deep Imaging Survey (DIS):  \citealt{martin09}, NIR J and K band: the UKIDSS Deep Extragalactic Survey (DXS) DR10 \citep{lawrence07}, MIR: IRAC instrument on board the Spitzer Space Telescope: SWIRE; \citealt{lonsdale03} and The Spitzer Extragalactic Representative Volume Survey (SERVS; \citealt{mauduit12}).  


\cite{mahony16}  presented the first LOFAR 150 MHz map of the LH with a sensitivity of 160 $\mu$Jy beam$^{-1}$ at a resolution of \asec{18.7} $\times$ \asec{16.4}. \cite{williams16}  presented the first LOFAR map of the Boo field at a resolution of \asec{5.6} $\times$ \asec{7.4} with an rms of 120 $\mu$Jy beam$^{-1}$. A deeper image of the Boo field, reaching an rms of 55 $\mu$Jy at its center, was presented by \citet{retanamontenegro18}. \cite{tasse20} (paper I) present the deepest, high-resolution (\asec{6}) low-frequency images and catalogues of the LH and Boo fields at 150~MHz and also describe the general method followed for the data reduction of the LoTSS Deep Fields. The even deeper LOFAR observations of the EN1 field are presented separately by \cite{sabater20} (paper II). 

One of the immediate science products of deep radio surveys is the determination of the radio source counts, which can provide useful comparison with counts predictions based on evolutionary models of radio source populations. In the present paper, we collectively exploit the LH, Boo and EN1 deep LOFAR data to derive the deepest radio source counts at 150~MHz ever. The derived source counts are compared with other existing determinations, as well as with state-of-the-art radio source evolutionary  models  (e.g. \citealt{wilman08}; \citealt{mancuso17}; \citealt{bonaldi19}). 

The outline of the paper is as follows. In Section 2 the data reduction and the imaging process followed to obtain the deep images of the LH, Boo and EN1 are described in brief. In Section 3, we summarize the source extraction process and we describe the derived source catalogues and corresponding properties. This is followed by an analysis of the source size distribution and of the catalogue incompleteness due to resolution bias (Section 4). Eddington bias and related incompleteness are discussed in Section 5. Section 6 presents the derived 150 MHz source counts and their comparison with state-of-the-art evolutionary models. We summarize our results in Section 7. Throughout this paper, we have used the convention $S_{\nu} \propto \nu^{\alpha}$.

\begin{figure*}[tb]
\begin{center}
\resizebox{0.45\hsize}{!}{
\includegraphics[angle=0]{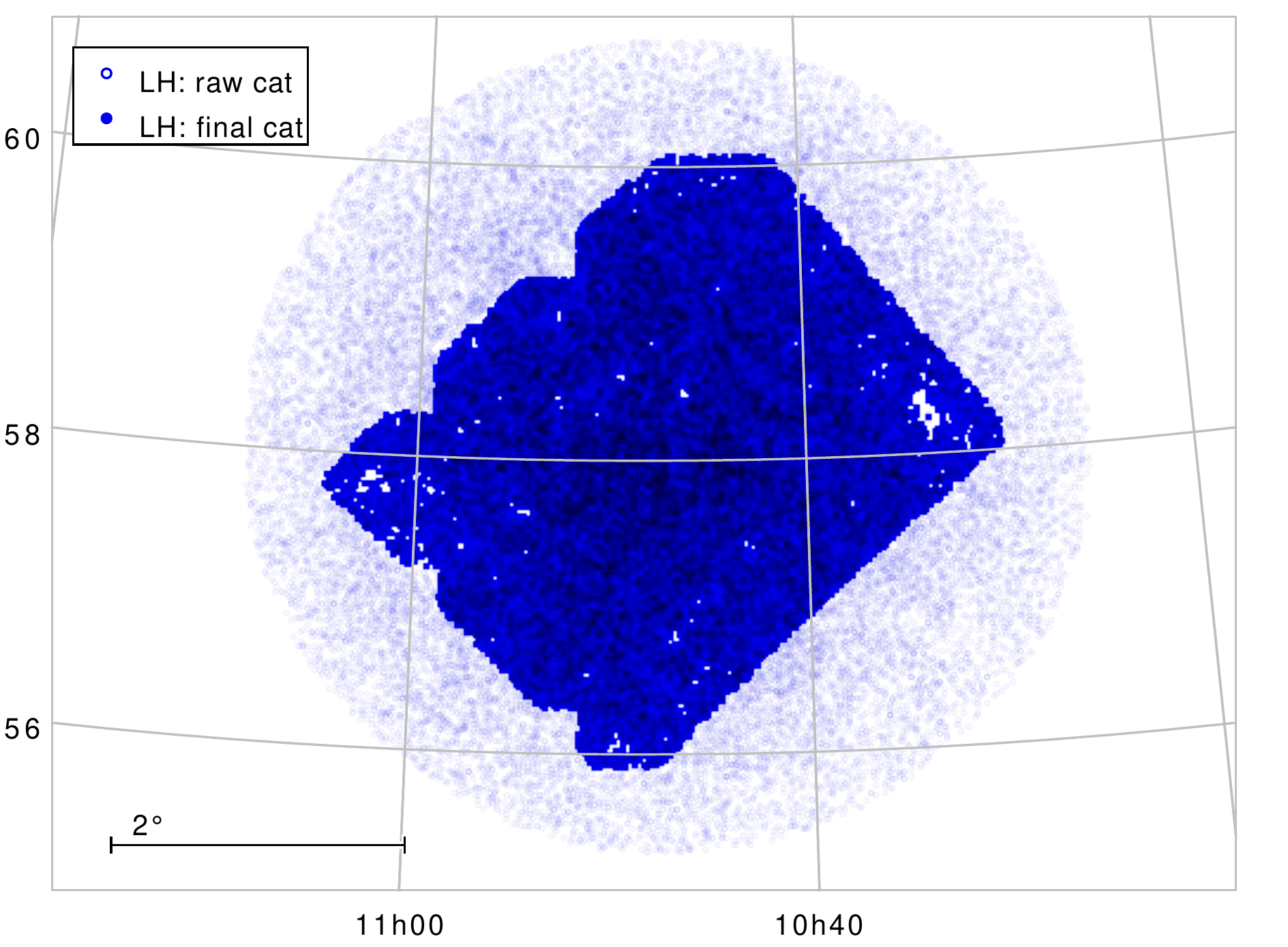}
}
\resizebox{0.45\hsize}{!}{
\includegraphics[angle=0]{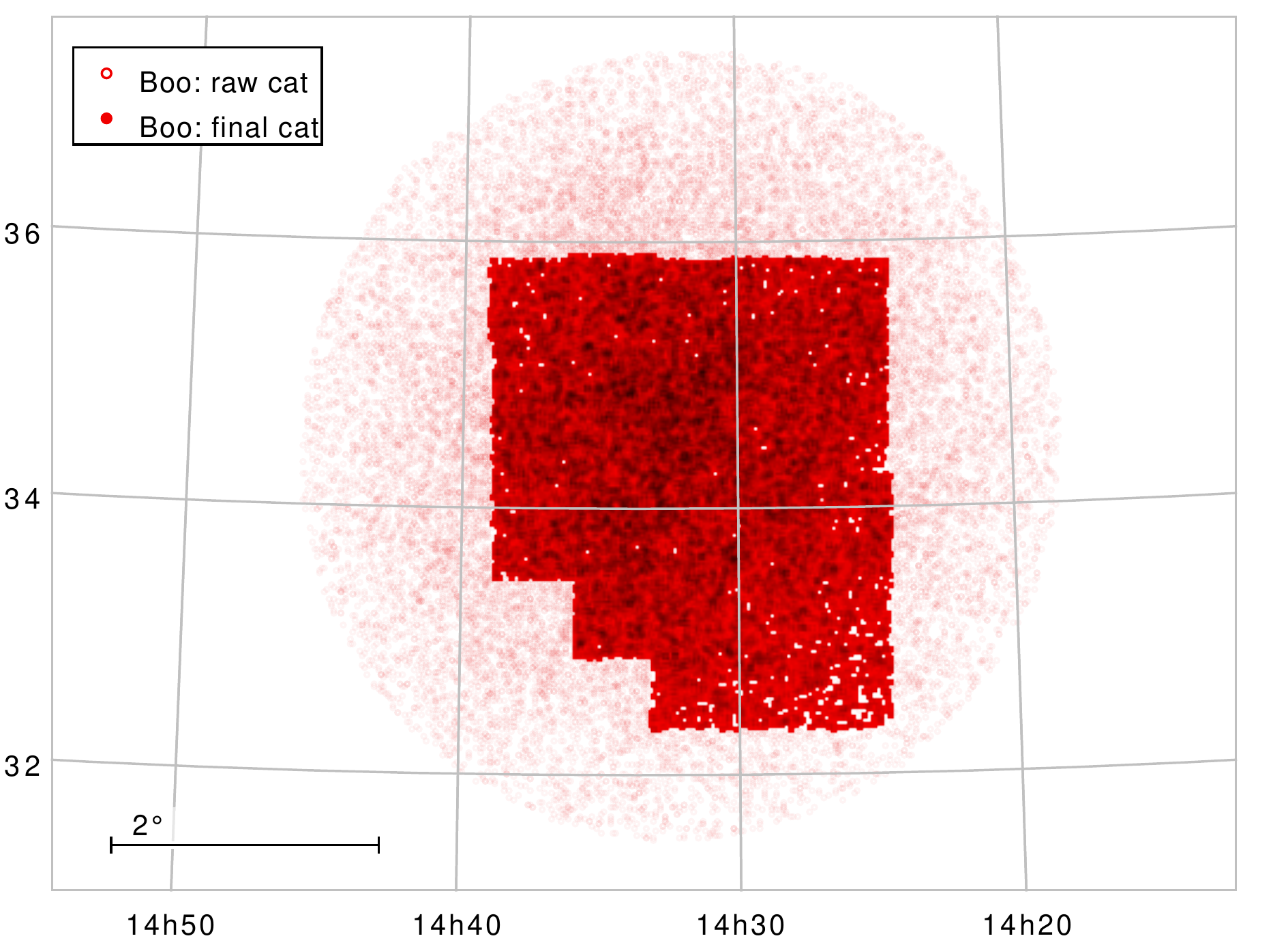}
}
\resizebox{0.45\hsize}{!}{
\includegraphics[angle=0]{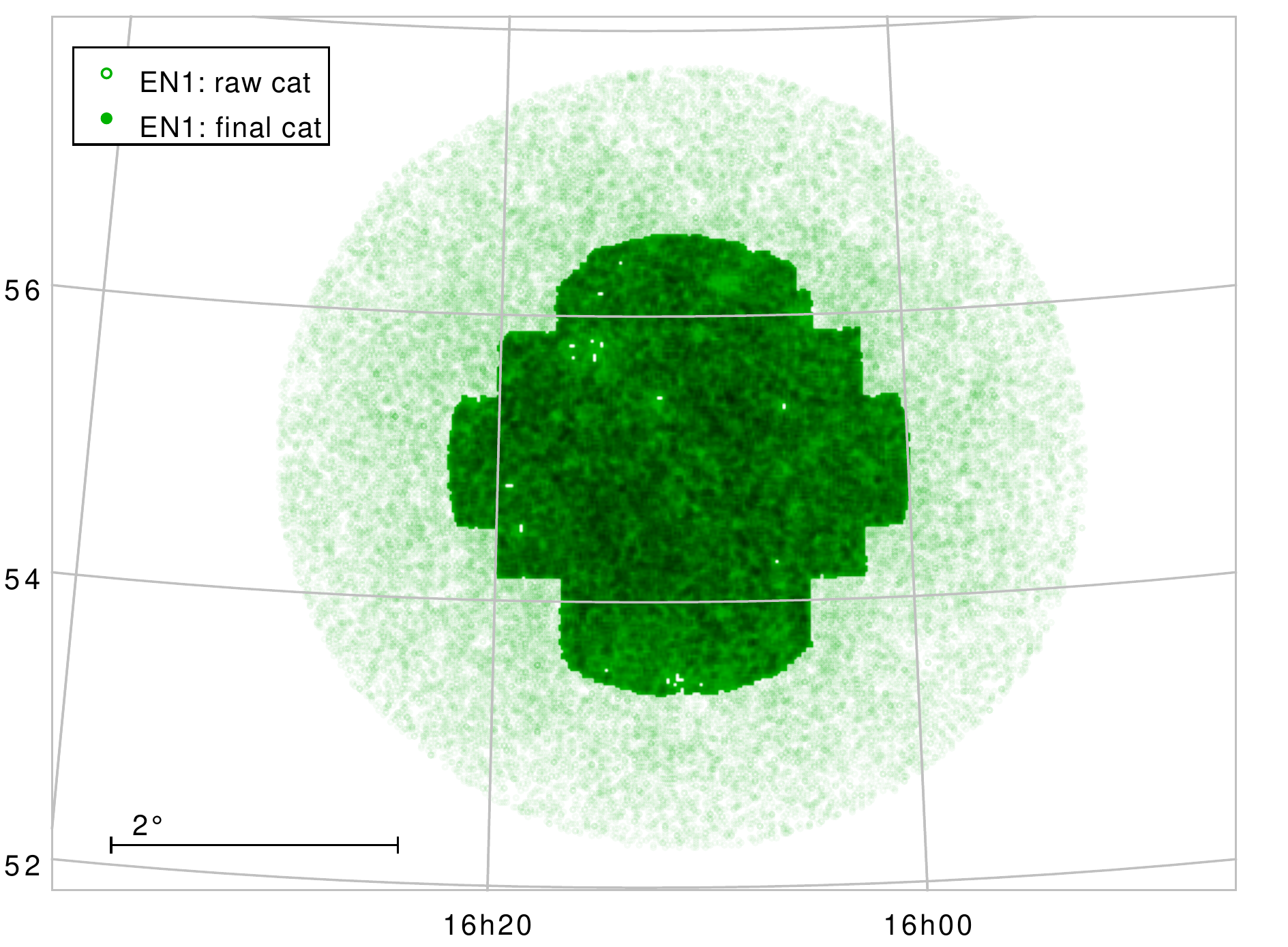}
}
\caption{LH (\textit{top left}), Boo (\textit{top right}) and EN1 (\textit{bottom}) fields targeted by LOFAR at 150 MHz. Light colors refer to the raw catalogues, cut at a  distance from the pointing center of  0.3 of the LOFAR 150 MHz  primary beam power. Darker colors refer to the final catalogues. The varying shape of their footprints highlights the regions with available optical/IR data. The areas of the optical/IR footprints are listed in Table~\ref{tab:image-properties}. 
}
\label{fig:deepfield}
\end{center}
\end{figure*}

\section{Observations and Data reduction}\label{sec:obs}

The observations and data reduction of the LoTSS Deep Fields are described in detail in paper I, but for completeness we provide a  brief summary below. 

Each of the deep fields was observed using the LOFAR High Band Antenna (HBA) in its $\texttt{HBA\_DUAL\_INNER}$ mode. Observations were taken in approximately 8hr blocks and the total integration times were 112, 80 and 164 hours for the LH, Boo and EN1 fields respectively\footnote{A full overview of the observation details is given in Table 1 of paper I (for the LH and the Boo fields) and in Table 1 of Paper II (for the EN1 field).}. The phase centers of the three pointings are listed in Table~\ref{tab:image-properties} (R.A., DEC.).  
The calibration of the data was completed in two steps. Firstly a direction independent calibration was performed using the \textsc{PreFactor} pipeline\footnote{https://github.com/lofar-astron/prefactor} which is  described in \cite{vanweeren16a} and \cite{williams16} and corrects for direction independent effects (see \citealt{gasperin19}). To efficiently deal with the large data rates, this pipeline is run on a compute cluster connected to the LOFAR archive (see \citealt{mechev18} and \citealt{drabent19}). The resulting data products are then calibrated with the latest version of \textsc{DDF-pipeline}\footnote{https://github.com/mhardcastle/ddf-pipeline} which is briefly outlined in Section 5.1 of \citealt{shimwell19} and detailed by paper I. This pipeline is based on the  kMS solver (\citealt{tasse14}; \citealt{smirnov15}) and the DDFacet imager (\citealt{tasse18}) to calibrate for direction-dependent effects, such as ionosphere-induced and beam model errors, and apply these solutions whilst imaging.

As described in \cite{tasse18},  for each deep field a single good observation is selected and run through \textsc{DDF-pipeline}. The resulting sky model, together with all observations from that particular field, are then input into a second run of \textsc{DDF-pipeline} which calibrates all the data off that sky model, before imaging all the data together and completing a final round of direction independent and direction-dependent self-calibration. The frequency coverage used to produce the images is 120 MHz to 168 MHz for Boo and LH and 115 MHz to 177 MHz for EN1\footnote{The exact central frequency of the imaged band is therefore 144 MHz for LH and Boo, and 146 MHz for EN1.}.

As described in papers I and II, the peak and integrated flux densities of the final images were rescaled by factors of 0.920, 0.859 and 0.796 for the LH, Boo and EN1 fields respectively. These scaling factors were derived from the comparison of the LOFAR flux densities with a variety of shallower radio surveys available at various frequencies over these fields. 
The minimum sensitivity reached at the center of the images (after rescaling) is $\sigma_c \sim $ 22, 32, 20 $\mu$Jy beam$^{-1}$, respectively, at a resolution of \asec{6} (see Table~\ref{tab:image-properties}). Although dynamic range effects are present around bright sources, in all cases the final image noise levels are within $\sim 10\%$ of the noise levels predicted from 8-hr depths, assuming an rms scaling with time $t^{-0.5}$. We note that the  noise measured in the Boo field is higher compared to the other two, also due to its lower declination.


\section{Source extraction, masking and deblending}\label{sec-debl}
Initial source catalogues were  extracted in each field using the PYthon Blob Detector and Source Finder (PyBDSF; \citealt{mohanrafferty15}.  The strategy followed for LH and Boo is detailed in paper I. In brief, the source detection threshold was set at 5$\sigma$ for the peak flux density and at 3$\sigma$ for the definition of the contiguous pixels used for the source Gaussian fitting, where $\sigma$ is defined as the local rms noise at the source position. To measure the background noise variations across the images, a sliding box of the size of 40 $\times$ 40 synthesized beams was used. For high signal-to-noise ($\geq$150) sources, the box size was reduced to 15$\times$15 synthesized beams in order to capture the increased local noise level more accurately. For EN1 a slightly different set of parameters was used (see Table C.1 of paper II). 
The PyBDSF wavelet decomposition mode was used in all fields to better describe complex sources characterized by very extended emission. PyBDSF produces  source catalogues, by grouping together Gaussian components that  belong to the same island of emission.  A flag is assigned to each source according to the number of Gaussian components fitted and grouped together to form a source: `S' and `M' refer to sources  fitted by a single and multiple Gaussian components respectively, whereas `C' means that the source lies within the same island as another source. For a more detailed description of the method and format of the catalogues, see the webpage\footnote{http://www.astron.nl/citt/pybdsf/} and \cite{shimwell19}. The catalogues were cut at a distance  from the pointing centre roughly corresponding to 0.3 of the 150 MHz LOFAR primary beam power (corresponding to fields of view of about 25 deg$^2$). The footprints of these initial PyBDSF catalogues (hereafter referred to as {\it raw} catalogues) are shown in light colors in Figure~\ref{fig:deepfield}. The total number of sources over these footprints is respectively $50,112$ (LH), $36,767$ (Boo) and $69,954$ (EN1).

Deep and wide optical and IR data are available for part of the LoTSS Deep Fields. Over these  sub-regions, we were able to carry out a further process of multi-wavelength cross-matching and source characterisation (see Paper III). This process, based on a combination of statistical  methods and visual classification schemes, allowed us to identify the host galaxies of over 97\% of the detected radio sources\footnote{97.6\% for EN1 and LH; 96.9\% for Boo.}. In addition it allowed us to produce a cleaner and more reliable radio source catalogue by a) removing spurious detections  (mainly artefacts introduced by residual phase errors around bright radio sources), and b) mitigating PyBDSF failures in correctly associating components to a source. Incorrect associations can occur in two main ways. Firstly, radio emission from physically distinct nearby sources could be associated as one PyBDSF source (blended sources). Such blends are more common at the faint end of the radio catalogues, where the source density is higher. Secondly, sources with multiple components could be incorrectly grouped into separate PyBDSF sources due to a lack of contiguous emission between the components. For example, this can occur for sources  with well separated double radio lobes.

An extensive description of the aforementioned process is given by paper III (see also \citealt{williams19}). Here we only provide a brief summary. All sources were evaluated through a decision tree to select those that require direct visual inspection and those that can be cross-matched through a Likelihood Ratio (LR) analysis. Sources with secure radio positions (f.i. those described by compact single Gaussian components) were selected as suitable for the LR method. Very extended, complex or multiple Gaussian sources were classified through visual inspection froma  group consensus, as well as sources that turn out to have an unreliable LR identification. Artifacts are generally found among sources with no reliable identification. Blended sources are typically recognized by the fact that (at least some of) their Gaussian components have very reliable distinct identifications. When confirmed through visual inspection, they are deblended.

After this cleaning process, the radio catalogues collect  respectively $31,163$ (LH), $19,179$ (Boo) and $31,645$ (EN1) sources,  distributed respectively over 10.3 deg$^2$ (LH), 8.6 deg$^2$ (Boo) and 6.7 deg$^2$ (EN1).  In the following we will refer to these deblended and associated catalogues as {\it final} catalogues. The footprints of the final  catalogues are shown in dark colors in Figure~\ref{fig:deepfield}. The irregular shape of these footprints follows the optical/IR sky coverage, corresponding to the region where source association and cross-identification is performed. We note that `holes' are present in such footprints, due to the fact that regions with optically bright stars (which typically produce artifacts in their surroundings) were masked.  

\begin{figure}[tb]
\begin{center}
\resizebox{0.9\hsize}{!}{
\includegraphics[angle=0]{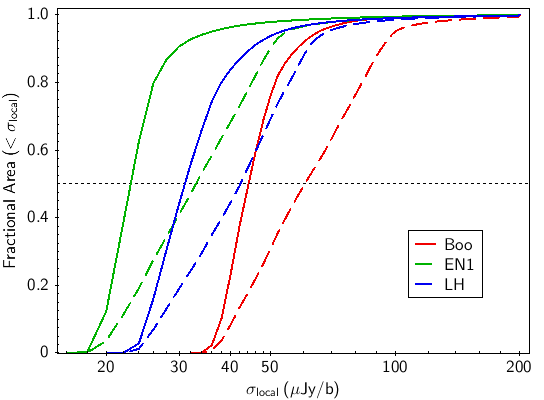}
}
\caption{Visibility functions of the raw (dashed lines) and final (solid lines) catalogues presented in this paper. Blue, red and green colors correspond to the LH, Boo and EN1 fields, respectively. The visibility functions represent the cumulative fraction of the total area of the noise map characterized by a noise lower than a given value. We caution that the total area covered by the final catalogues is much smaller than the one covered by the raw catalogues (see Tab. \ref{tab:image-properties}).}
\label{fig:noisehist}
\end{center}
\end{figure}

In addition we have generated pixel-matched images in each waveband and extracted forced aperture-matched photometry from ultraviolet to infrared wavelength (paper III), deriving high-quality photometric redshifts for around 5 million objects across the three fields (see Duncan et al., 2020, paper IV of this series, for more details). The raw and final radio catalogues, as well as the optical/IR and photometric catalogues, are available on the LOFAR Surveys Data Release site web-page\footnote{http://www.lofar-surveys/releases.html}.

\subsection{Visibility function of raw and final catalogues}
\label{section:noise-characteristics}
Figure \ref{fig:noisehist} shows the so-called visibility function (i.e. the cumulative fraction of the total area of the noise map characterized by noise measurements lower than a given value) for the LH (blue), Boo (red) and EN1 (green) fields. Raw and final catalogues are indicated respectively by the dashed and solid lines. We note that the visibility functions of final catalogues are significantly steeper than those of the raw catalogues. This is due to the fact that  the final catalogues are mostly confined in the inner, most sensitive parts  of the LOFAR fields. As a consequence the median noise is significantly lower for final than for raw catalogues (see Table~\ref{tab:image-properties}).


\begin{table}[t]
    \caption{Parameters describing the unresolved/resolved sources' dividing lines (see Eqs.~\ref{eqn:envelope} and \ref{eqn:envelope2}) for the LH, Boo and EN1 catalogues.}
   \label{tab:flux-scale}
    \centering
   \begin{tabular}{c|cccc}
   \hline
   
      Field &  A & B & \multicolumn{2}{c}{$\%_{\rm resolved}$} \\
        & & & raw & final\\
         \hline
           & &  & &  \\
      LH   & 1.15 & 3.0 & 34 & 25 \\
      Boo  & 1.00 & 2.0 & 47 & 38 \\
      EN1  & 1.07 & 3.0 & 35 & 24 \\
      \hline
    \end{tabular}
\end{table}

\begin{figure}[!th]
\begin{center}
\resizebox{0.99\hsize}{!}{
\includegraphics[angle=0]{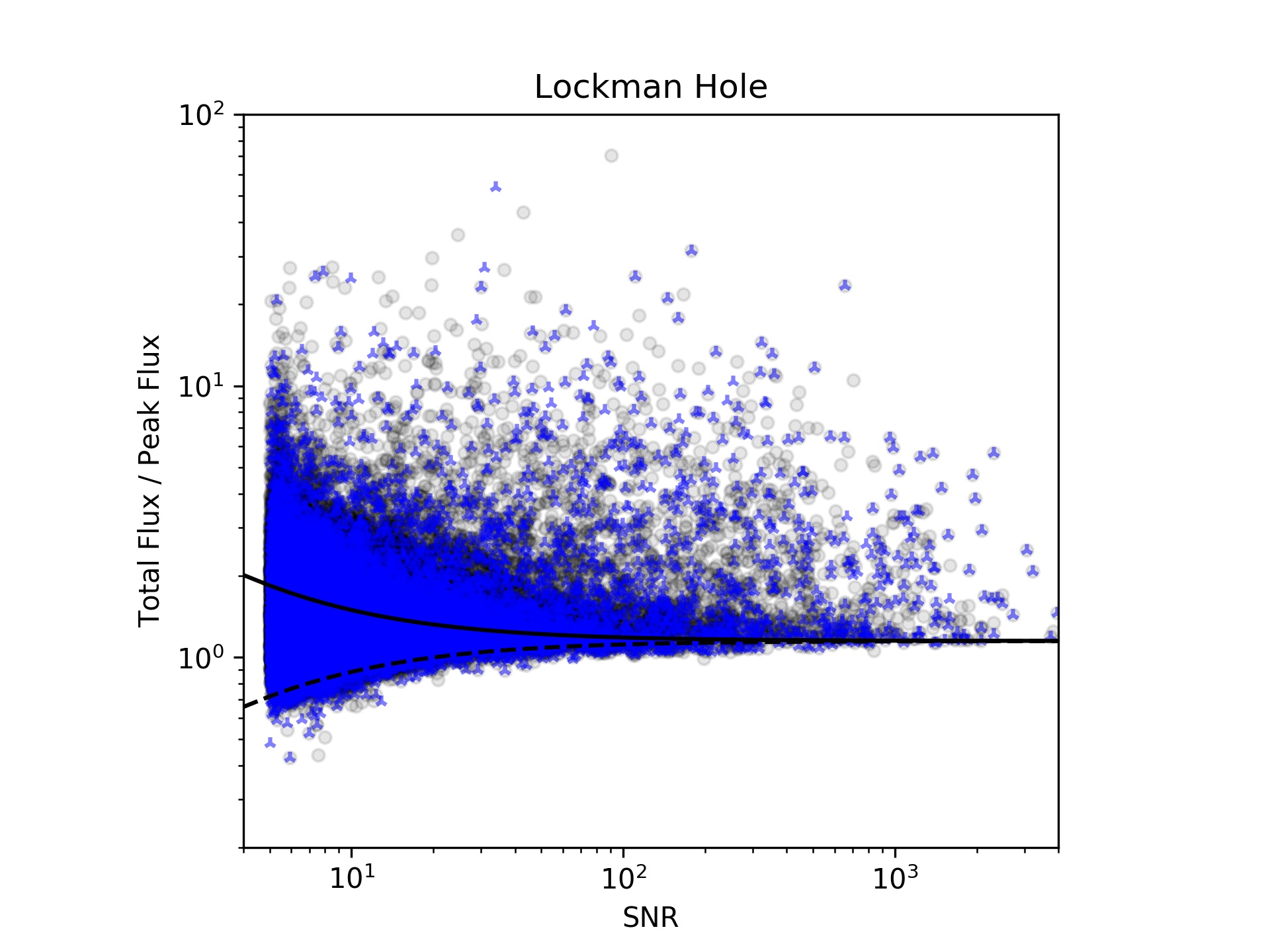}
}
\resizebox{0.99\hsize}{!}{
\includegraphics[angle=0]{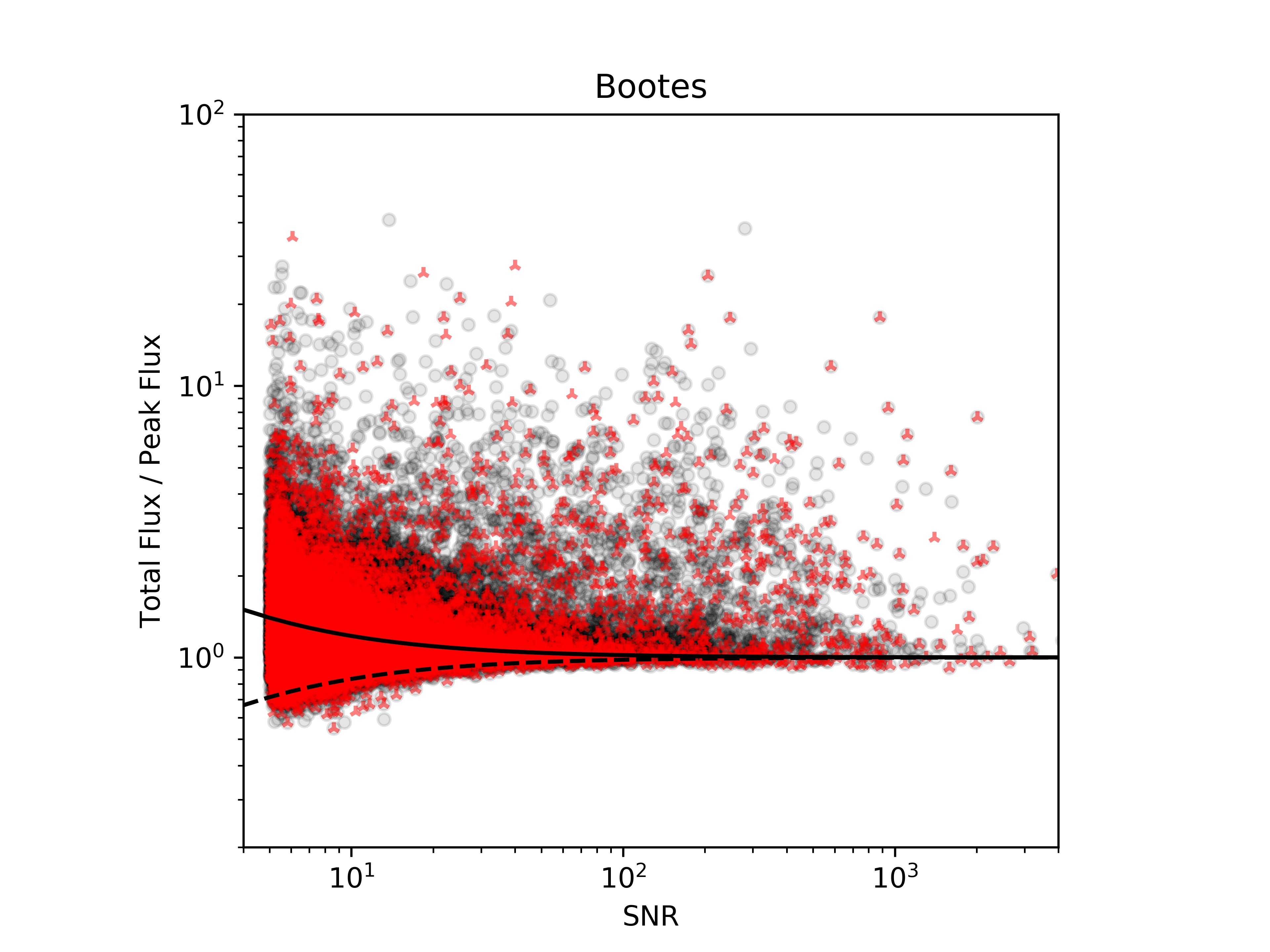}
}
\resizebox{0.99\hsize}{!}{
\includegraphics[angle=0]{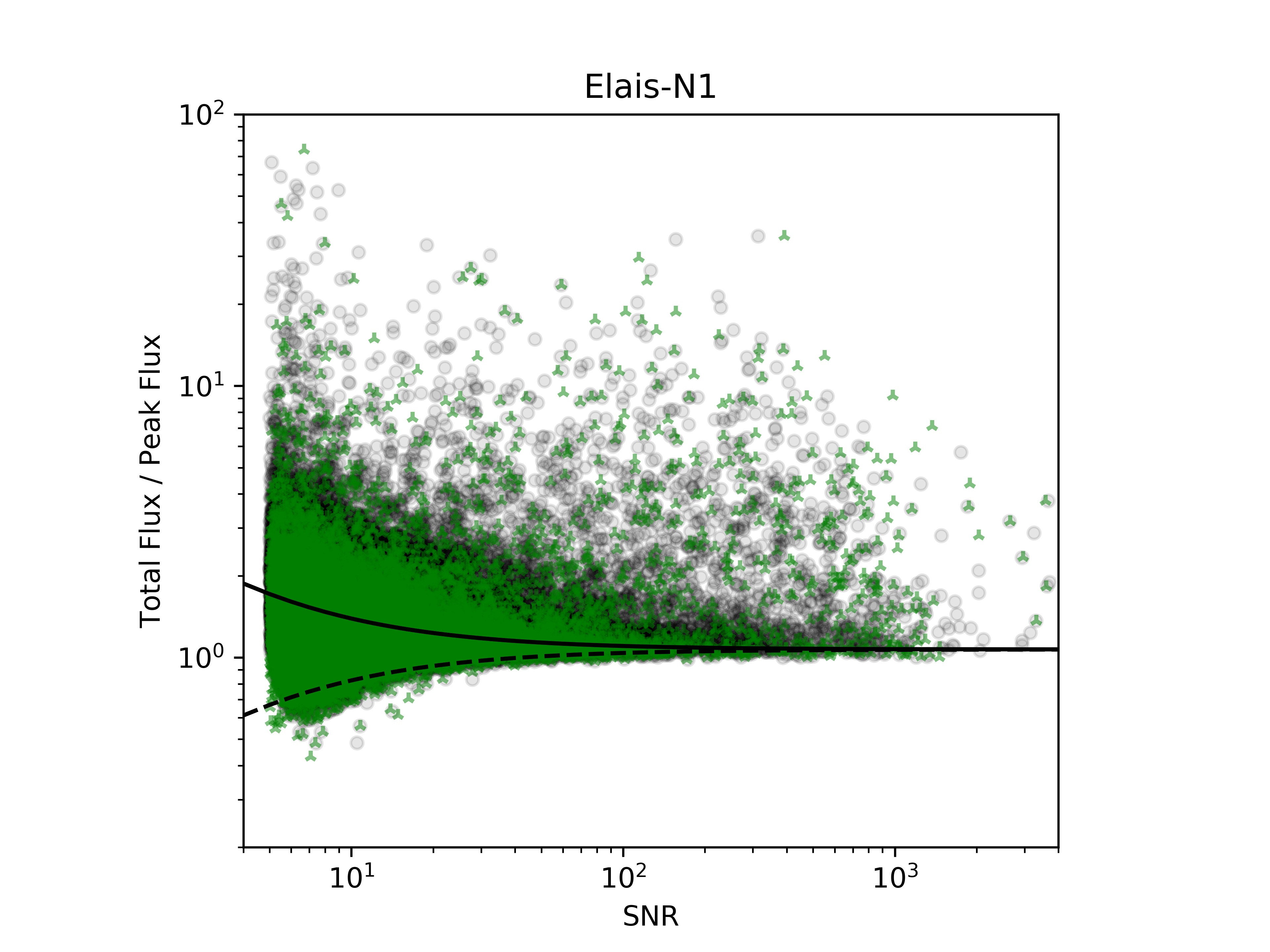}
}
\caption{Total to peak flux density ratio as a function of signal to noise ratio (SNR = $S_{\rm{peak}}/\sigma$) for both the raw (black transparent circles) and final ($\Yup$ symbols in blue, red and green colors) catalogues in the LH, Boo and EN1 fields (respectively from \textit{top} to \textit{bottom} panel). The dashed and solid lines represent the unresolved source distribution lower and upper envelopes respectively (see text for more details). }
\label{fig:envelope}
\end{center}
\end{figure}

\subsection{Source Size Deconvolution}
\label{sec:source-size}
Characterisation of resolved versus unresolved sources in our catalogues is important in order to correct the catalogues for the incompleteness introduced by so-called \textit{resolution bias} (described in Section \ref{sec:resolution-bias}). 
The total flux density ($S_{\rm{total}}$) of a source can be written as:
\begin{equation}
S_{\rm{total}}/S_{\rm{peak}} = \theta_{\rm{maj}}\theta_{\rm{min}}/b_{\rm{min}}b_{\rm{maj}}
\label{eqn:res-bias}
\end{equation}
where $S_{\rm{peak}}$ is the source peak flux density, $\theta_{\rm{min}}$ and $\theta_{\rm{maj}}$ are the source full-width-half-maximum (FWHM) axes, and $b_{\rm{min}}$ and $b_{\rm{maj}}$ are the restoring beam FWHM axes. In an ideal image, in the absence of noise,  the total flux density of a point source is equal to its peak flux density. 
In real images both the total and peak flux density measurements 
of point sources are affected by errors. 
This means that not all sources with $S_{\rm{total}}>S_{\rm{peak}}$ would be genuinely resolved sources. The $S_{\rm{total}}/S_{\rm{peak}}$ ratio as a function of signal-to-noise ratio (SNR = $S_{\rm{peak}}/\sigma$, where $\sigma$ is the local rms noise), can be used to establish a statistical criterion to establish if a source is likely extended or point-like (see e.g. \citealt{prandoni00b,prandoni06}). In Figure \ref{fig:envelope}, the ratio of the total to peak flux densities is shown as a function of SNR for both raw and final catalogues. 
A lower envelope of the source distribution can be defined by the following equation: 
\begin{equation}
S_{\rm{total}}/S_{\rm{peak}} = \rm{A} / (1+ \rm{B} /SNR)
\label{eqn:envelope}
\end{equation} 
where A and B are two free parameters (see dashed lines in each panel of Fig.~\ref{fig:envelope}). As expected, going to higher SNR, measurement errors get smaller.  At SNR$\gg$100 the 2nd term of Eq.~\ref{eqn:envelope} can be neglected, and the $S_{\rm{total}}/S_{\rm{peak}}$ tends to A. 
In an ideal case, where radial smearing is taken care of, the ratio of the total over the peak flux density for point sources should converge to a value of A=1 at very high SNRs. The DDFacet pipeline implements a facet dependent PSF which, for deconvolved sources, accounts for the impact of time and bandwidth smearing (\citealt{tasse14}). However, due to imperfect calibration of the PSF across the field and/or smearing of sources due to ionospheric distortions, the value of the ratio at high signal-to-noise sources can be found to be higher than 1 and can be field-dependent (as ionospheric effects are time and spatially dependent). The values of A for the LH, Boo and EN1 field are respectively 1.15, 1.00 and 1.07 (see Table~\ref{tab:flux-scale}). 
This could potentially mean that the Boo field is less affected by ionospheric smearing when compared with LH and EN1. The B value also changes depending on the field, with Boo showing a lower value than LH and EN1 (see Table~\ref{tab:flux-scale}), again indicating smaller errors in the determination of source flux densities. We notice that the parameters given in Table~\ref{tab:flux-scale} provides a good description of both raw and final catalogues. The lower envelopes can then be mirrored  around  the $S_{\rm{total}}$/$S_{\rm{peak}}$ = A axis to get the upper envelopes: 
\begin{equation}
S_{\rm{total}}/S_{\rm{peak}} = \rm{A} \cdot (1+ \rm{B} /SNR)
\label{eqn:envelope2}
\end{equation} 
Sources lying above the upper envelopes (dashed black lines in each panel) are then considered to be truly extended or resolved sources. Sources below the upper envelopes are considered to be point sources. The fraction of resolved sources in each field is given in Table~\ref{tab:flux-scale}. In final catalogues the fraction of resolved sources vary from  24-25\% (EN1 and LH) to 38\% (Boo). The $\sim 10\%$ higher fractions observed in raw catalogues reflect the larger number of bright extended sources detected in their larger FoV. These fractions should be considered as indicative, as they depend on the criteria used to define them. \cite{sabater20}, for instance, as part of their detailed analysis of the EN1 field, used more stringent criteria, which also include additional sources of errors for the source flux densities, and estimated that between 4 and 11\% of the sources in the EN1 raw catalogues are genuinely extended (see paper II for more details). Nevertheless, we decided to apply the same approach to all fields, and to both final and raw catalogues, to enable a consistent statistical analysis of the source size distribution in the three fields (see Sect.~\ref{sec:resolution-bias}).


\section{Source Size Distribution and Resolution Bias}
\label{sec:resolution-bias}

In deriving the source counts, the completeness of the catalogues in terms of total flux density needs to be estimated. Such completeness depends on source angular sizes, since, as shown by Eq.~\ref{eqn:res-bias}, a larger source of a given total flux density will drop below the 5$\sigma$ limit of a survey more easily than a smaller source of the same total flux density. This effect, called  \textit{resolution bias}, results from the fact that the detection of a source depends on its peak flux density. Following \citet{prandoni01,prandoni06}, we can use Eq.~\ref{eqn:res-bias} to calculate the approximate maximum deconvolved size ($\Theta_{\rm{max}}$) a source of a given total flux density, $S_{\rm{total}}$, can have before dropping below the 5$\sigma$ limit of the catalogue:
\begin{equation}
\Theta_{\rm{max}}=\Theta_N \sqrt{(S_{\rm{total}}/(5\sigma)-1}
\label{eq:thetamax}
\end{equation} 
where $\Theta_{\rm{N}}$ $\equiv$ $\sqrt{b_{\rm{maj}}b_{\rm{min}}}$ is the geometric mean of the restoring beam axes. In our case $\Theta_{\rm{N}}$=$b_{\rm{maj}}$=$b_{\rm{min}}$=\asec{6}. 

\begin{figure*}[!t]
\begin{center}
\vspace{-1cm}
\resizebox{0.42\hsize}{!}{
\includegraphics[angle=0,clip]{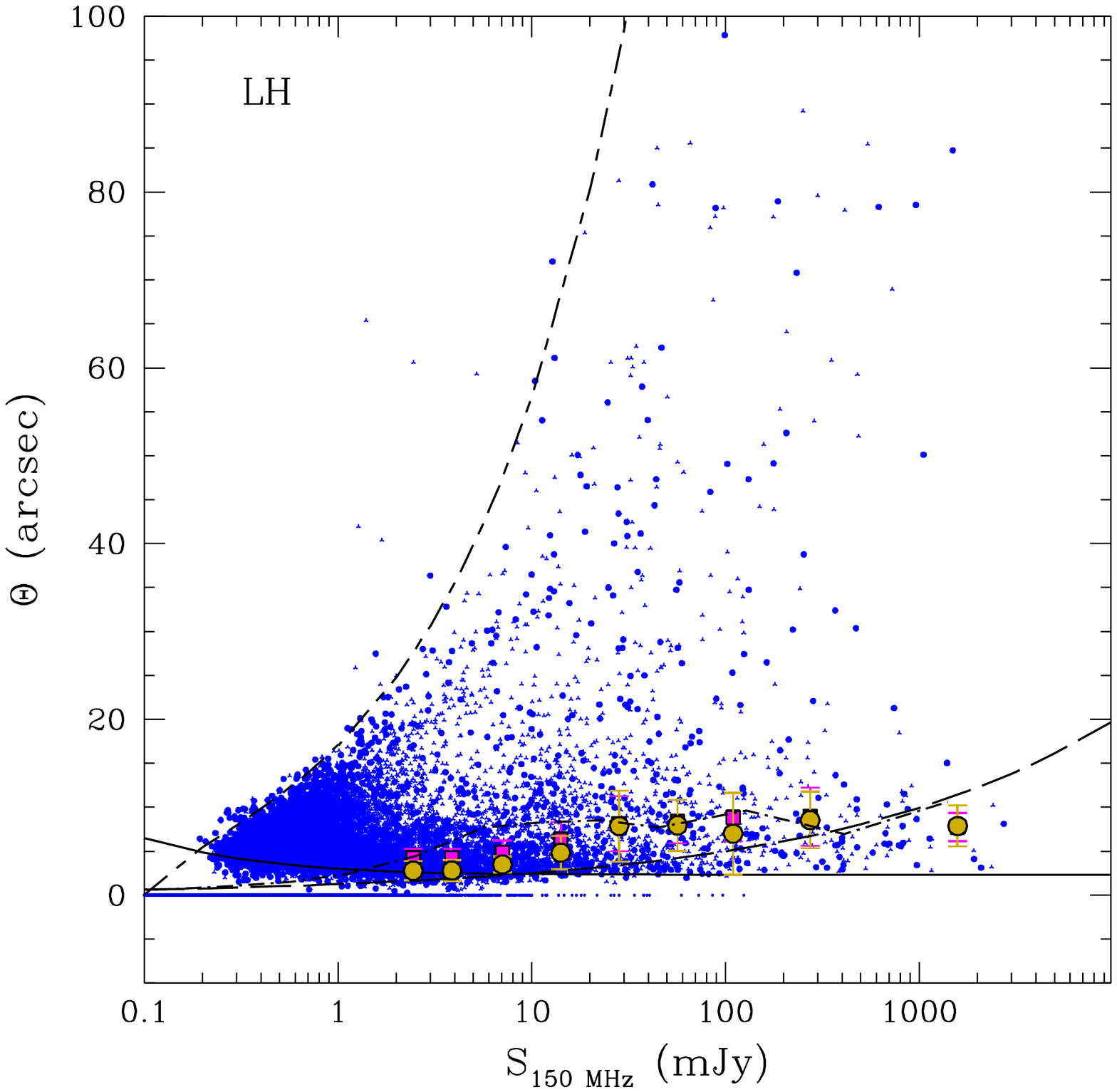}
}
\resizebox{0.42\hsize}{!}{
\includegraphics[angle=0,clip]{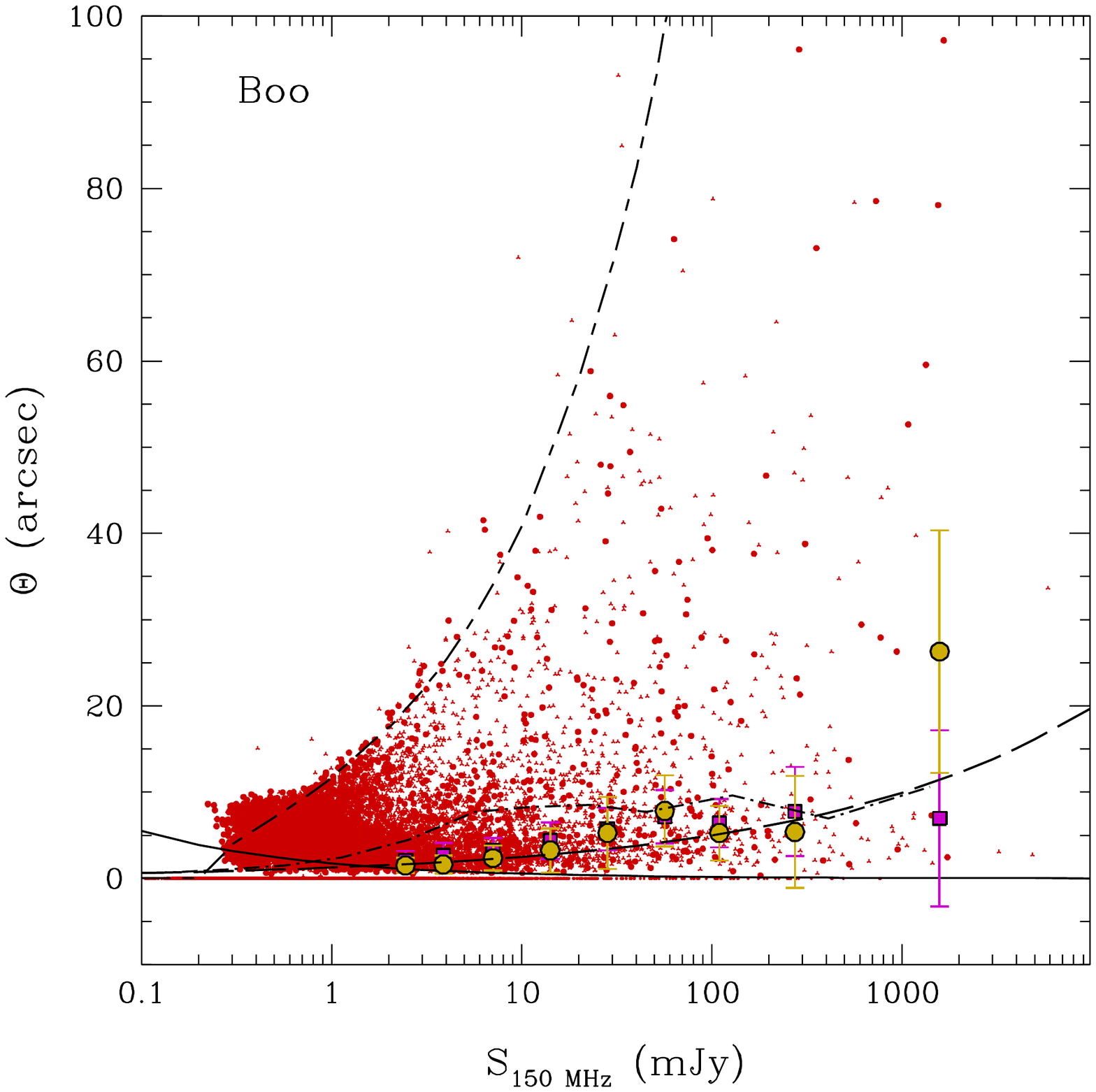}
}
\vspace{-2.5cm}

\resizebox{0.42\hsize}{!}{
\includegraphics[angle=0,clip]{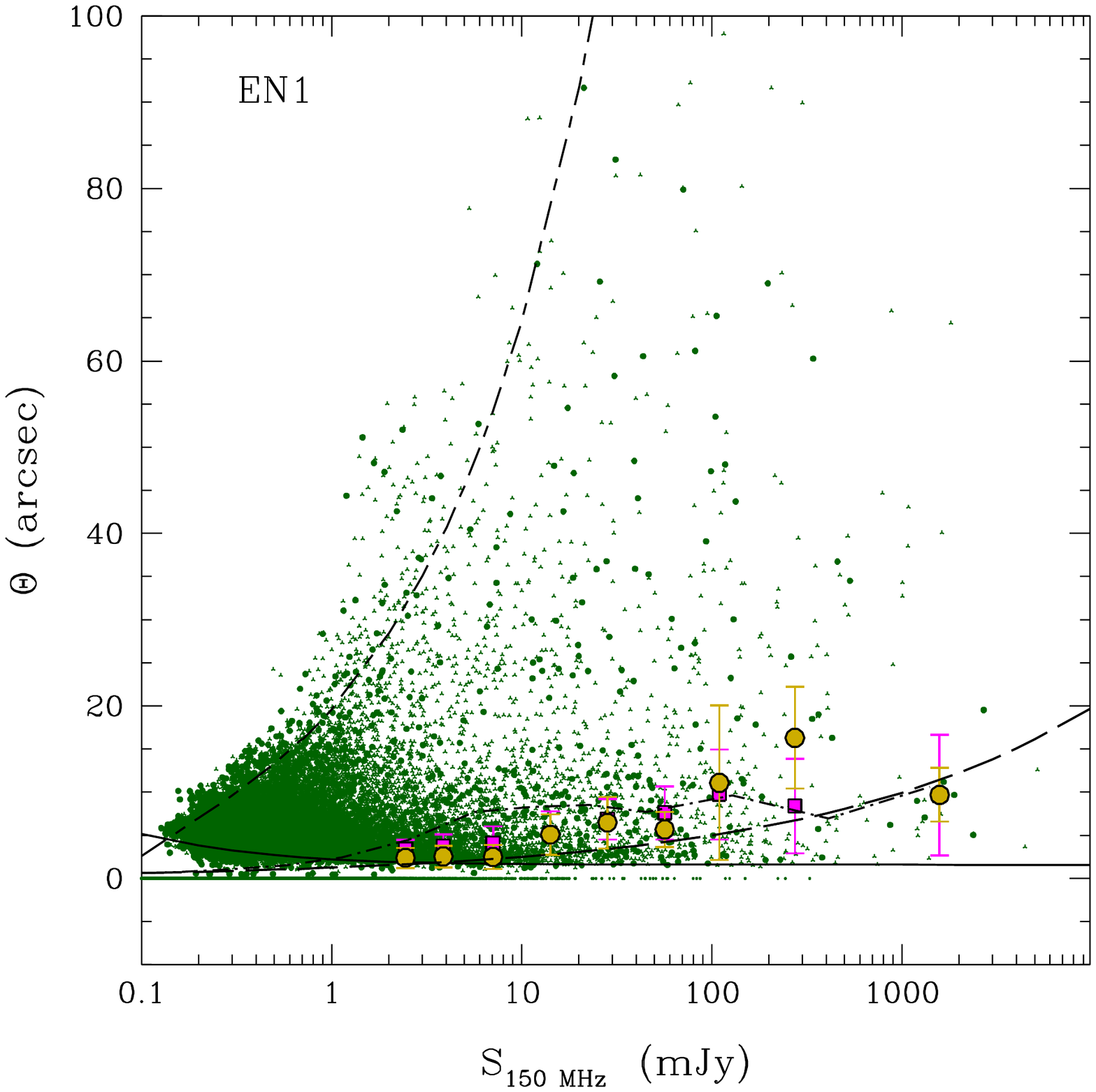}
}
\vspace{-1.5cm}
\caption[]{Source intrinsic (deconvolved) angular sizes as a function of the measured 150 MHz total flux densities. Deconvolved sizes are defined as the geometric mean of the major and minor FWHM axes, except for well resolved radiogalaxies, which are better described by their major axis. Deconvolved sizes of point sources are set to zero. Raw ($\centerdot$) and final ($\Yup$) catalogues of the LH, Boo and EN1 fields are shown in the \textit{top left}, \textit{top right} and \textit{bottom} panels, respectively. The short-long-dashed lines in the three panels define the maximum size ($\Theta_{\rm{max}}$) a source can have for a given measured total flux density before dropping below the detection threshold. The solid lines give the minimum size ($\Theta_{\rm{min}}$) below which deconvolution is not considered reliable. Both lines have been drawn assuming the median noise in the masked area (see last column of Table \ref{tab:image-properties}). The long-dashed lines indicate the \citet{windhorst90} median size - flux density relation, converted to 150 MHz, while the dot-dashed lines indicate the median size - 150 MHz flux density relation derived from the simulated T-RECS catalogues directly at 150 MHz (\citealt{bonaldi19}). The filled black-bordered magenta squares and golden circles with error bars represent the median source sizes for the raw and final catalogues respectively. Medians are computed only for those flux density bins where unresolved sources represent less than 50\% of the total number of sources.
\label{fig:size-flux}}
\end{center}
\end{figure*}

In Figure \ref{fig:size-flux} we show the deconvolved source sizes as a function of the total flux density for both raw and final catalogues. Each panel corresponds to a different field: LH (top-left), Boo (top-right) and EN1 (bottom).  Deconvolved sizes are defined as the geometric mean of the major and minor FWHM axes, except for well resolved radio galaxies, which are better described by their major axis. Deconvolved sizes of point sources are set to zero. As expected, the upper envelope of the source size distributions approximately follow the $\Theta_{\rm{max}} - S_{\rm{total}}$ relation (short-long-dashed line) in all fields.

Equations \ref{eqn:res-bias} and \ref{eqn:envelope2} can also be used to derive an approximate minimum intrinsic angular size ($\Theta_{\rm{min}}$) that can be resolved reliably as a function of the source peak flux density:
\begin{equation}
\Theta_{\rm{min}}=\Theta_{\rm{N}} \sqrt{\rm{A\cdot (1+B/SNR})-1} \; .
\label{eq:thetamin}
\end{equation}
The curve representing $\Theta_{\rm{min}}$ is shown in Figure \ref{fig:size-flux} by the solid lines. 

In order to quantify the fraction of sources larger than $\Theta_{\rm{max}}$, and in turn the incompleteness affecting our catalogue, we need to know the true intrinsic radio source size distribution within the flux density range probed by our survey. We start assuming the empirical integral distribution proposed by \cite{windhorst90} for  1.4 GHz-selected samples:
\begin{equation}
    h(>\Theta) = {\rm{exp}}[-\ln{2}\, (\Theta/\Theta_{\rm{med}})^{q}]
    \label{eq:cumdistr}
\end{equation}
where $q=0.62$ and the median source size varies with the total flux density as follows:
\begin{equation}\label{eq-sizew90}
    \Theta_{\rm{med}} = k \times  (S_{\rm{1.4GHz}})^{m}
\end{equation}
with $k$=\asec{2}, $m=0.3$, $S_{\rm{1.4GHz}}$ expressed in mJy. The \citet{windhorst90} relations are extensively used in the literature to estimate the resolution bias, either for 1.4 GHz selected samples (see e.g. \citealt{prandoni01,prandoni18}; \citealt{huynh05}; \citealt{hales14a}), or for surveys at other frequencies, including LOFAR HBA ones (\citealt{mahony16}; \citealt{williams16}; \citealt{retanamontenegro18}).
We converted the median size - flux density relation to 150 MHz assuming a spectral index $\alpha=-0.7$. This assumption is appropriate for radio catalogues dominated by faint sub-mJy radio sources. Indeed  spectral index analyses performed using shallower ($S_{\rm 150\; MHz}\ga 1$ mJy) LOFAR observations of the Bo$\ddot{\rm o}$tes and LH fields, report overall median spectral index values of $\alpha_{\rm 150\, MHz}^{\rm 1.4\, GHz}= -0.73\pm 0.33$ and $-0.78\pm 0.24$, for AGN and star-forming galaxies respectively (Bo\"otes; \citealt{calistro-rivera17}), as well as a flattening of the spectral index  going to lower flux densities, with a median value of $\alpha_{\rm 150 MHz}^{\rm 1.4 GHz}=-0.7_{-0.04}^{+0.02}$ at $S_{\rm 150\, MHz}\sim 1-2$ mJy (LH; \citealt{mahony16}). 

As shown in Figure~\ref{fig:size-flux} the median sizes of both raw and final catalogues (respectively indicated by filled black-bordered magenta squares and golden circles with error bars) are compared with the \citet{windhorst90} size - flux density relation converted to 150 MHz (long-dashed line). We see a discrepancy at intermediate flux densities ($10 - 100$ mJy), where the measured sizes appear in slight excess to what was predicted by \citet{windhorst90}. We therefore decided to consider also the median size -- flux density relation derived from the Tiered Radio Extragalactic Continuum Simulation (T-RECS) catalogues at 150 MHz (\citealt[dot-dashed line]{bonaldi19}), which implement different size -- flux density scaling relations for star-forming galaxies and AGN. This seems to better reproduce our measured sizes at flux densities $S_{\rm 150 MHz}\sim 10-100$ mJy, where extended radio galaxies (with typical sizes of hundreds of kpc) are expected to provide a significant contribution to the total radio source population. We note, however, that the afore-mentioned analysis is limited to flux densities $S_{\rm 150\; MHz} \ga 2$ mJy, while the large majority of the sources in the LoTSS Deep Fields are fainter. Most of these sources cannot be reliably deconvolved, implying that no direct information on their size distribution can be obtained. Several attempts have been made to estimate the intrinsic source sizes at sub-mJy flux densities, based on deep samples carried out over a wide range of observing frequencies (from 330 MHz to 10 GHz). Some of these works have proposed a steepening of the \citet{windhorst90} median size - flux density relation at sub-mJy fluxes, with 
$m=0.4-0.5$ in the range $0.1-1$ mJy (\citealt{richards00}; \citealt{bondi03,bondi08}; \citealt{smolcic17}). A smooth transition from a flatter to a steeper relation at sub-mJy flux densities could again be justified by a smooth transition from a flux regime dominated by extended radio galaxies ($S\gg1$ mJy) to a flux regime dominated by  radio sources triggered by star-formation (or by composite SF/AGN emission), confined within the host galaxy. 

\begin{figure}[!t]
\begin{center}
\vspace{-1cm}
\resizebox{1.00\hsize}{!}{
\includegraphics[angle=0]{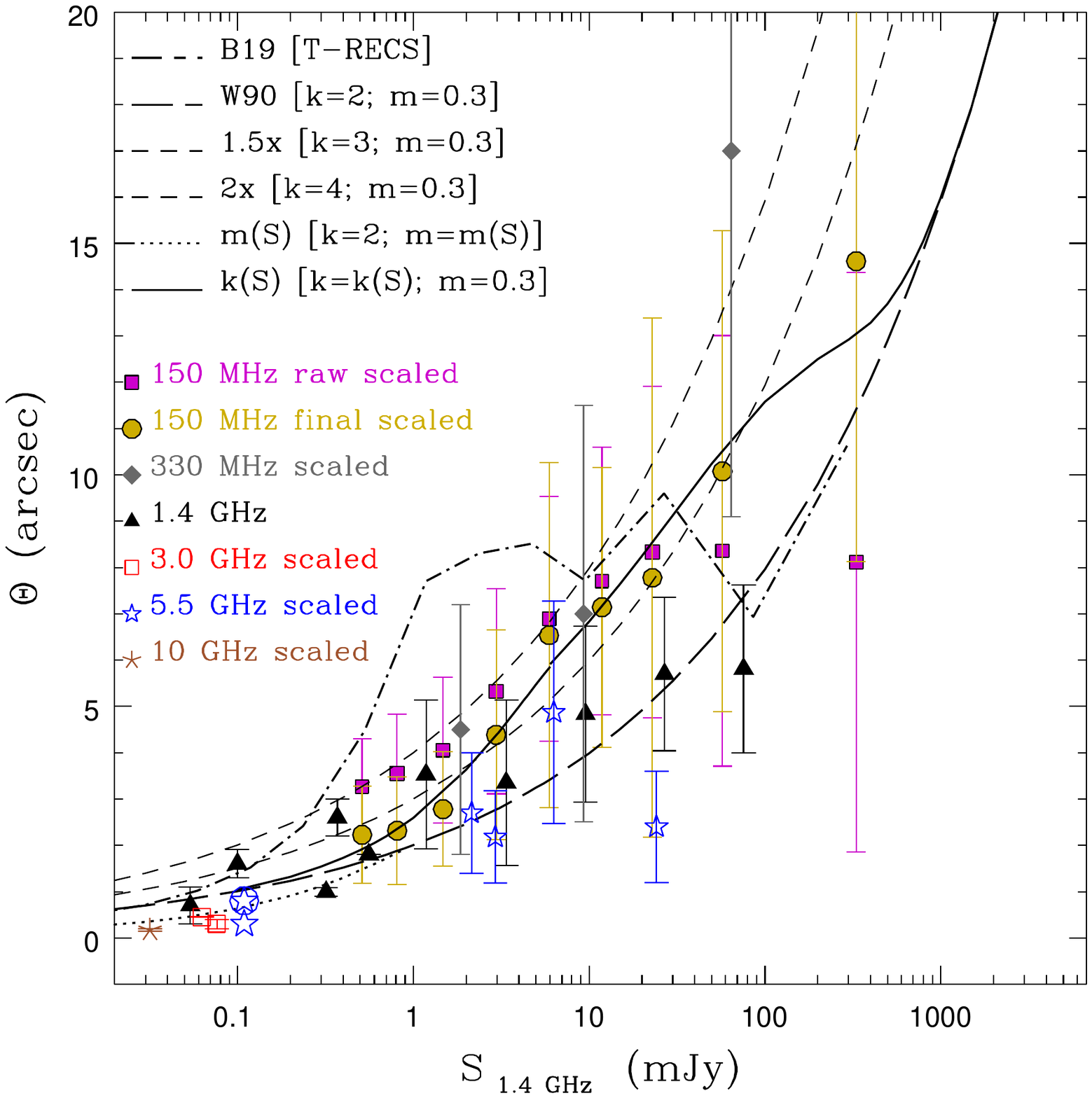}
}
\vspace{-2.5cm}
\caption[]{Source median angular size vs. 1.4 GHz total flux density, as estimated in some of the deepest radio samples available so far. Different colors/symbols correspond to different observing frequencies: 330 MHz (grey filled diamonds -  \citealt{owen09}); 1.4 GHz (black filled triangles -  \citealt{richards00}; \citealt{bondi03,bondi08}; \citealt{muxlow05}; \citealt{prandoni18}); 3 GHz (red empty squares - \citealt{bondi18}; \citealt{cotton18}); 5.5 GHz (blue stars - \citealt{prandoni06}; \citealt{guidetti17}); 10 GHz (brown asteriscs - \citealt{murphy17}). Also shown are the median sizes measured in our raw and final catalogues (150 MHz), combined together (filled black-bordered magenta squares and golden circles). We note that \citet{guidetti17} gives different median sizes for the AGN and star-forming galaxy sub-populations. The latter population is indicated as a circled blue star in the figure. All flux densities have been converted to 1.4 GHz, assuming a spectral index $\alpha= -0.7$. Various median size -- flux density relations are shown for comparison: the ones proposed by \citet{bonaldi19} and \citet{windhorst90} (dot-dashed and long-dashed lines respectively), and some revised versions of the latter. The short dashed lines show the relations obtained by rescaling the \citet{windhorst90} relation by 1.5$\times$ and 2$\times$ (i.e. assuming $k=3$ and $k=4$ in Eq.~\ref{eq-sizew90}); the dotted line assumes a smooth transition between $m=0.3$ and $m=0.5$ going from mJy to sub-mJy flux densities,a s described by Eq.~\ref{eq:revw90}; the solid line assumes a value of $k$ varying with flux density according to Eq.~\ref{eq:w90ks} (see text for more details). 
\label{fig:median-size}}
\end{center}
\end{figure}

In order to establish which size--flux density relation would best quantify the incompleteness of our catalogues we have decided to include in our analysis the results from other deep surveys. Figure~\ref{fig:median-size} shows the existing measurements of (median) source sizes in various flux density bins for a number of surveys (different colors/symbols refer to different observing frequencies). Also shown are the median sizes derived by combining together the three LoTSS fields (raw and final catalogues, respectively indicated by filled black-bordered magenta squares and golden circles). To make the comparison meaningful, all flux densities referring to a different observing frequency have been converted to 1.4 GHz, assuming $\alpha=-0.7$.  Also shown are 
various size -- flux density relations: the ones proposed by \citet[converted to 1.4 GHz]{bonaldi19} and \citet{windhorst90} (dot-dashed and long-dashed lines respectively), and some modifications of the latter. The short dashed lines show the ones obtained by rescaling the \citet{windhorst90} relation by 1.5$\times$ and 2$\times$ (i.e. assuming k=3 and k=4 in Eq.~\ref{eq-sizew90}), while the dotted line  assumes a smooth transition between $m=0.3$ and $m=0.5$ going from mJy to sub-mJy flux densities, i.e.:
\begin{equation}
    m=m(S)=0.3+0.2 \times \rm{exp}(-S_{\rm{1.4GHz}}^2)
        \label{eq:revw90}
\end{equation}
with $S_{\rm{1.4GHz}}$ expressed in mJy. Focusing on the sub-mJy regime, it is clear that both the \citet{windhorst90} and the steeper $m(S)$ relations are consistent with the observed sizes, especially when considering only the 1.4 GHz surveys (black filled triangles). Surveys undertaken at higher frequencies seem to point towards the steeper relation, but these samples may be biased towards a flatter spectrum population, resulting in an over-estimation of the flux densities once converted to 1.4 GHz assuming a too steep spectral index. We also caution that higher frequency surveys more easily miss extended flux, and resolution bias issues can  indeed mimic a steepening of source median sizes getting close to the flux limit of a radio survey.  At larger flux densities ($S_{\rm{1.4GHz}} \ga 1$ mJy) the median sizes are observed to lie between the \citet{windhorst90} relations described by $k=2$ and $k=4$, with a tendency for larger sizes going to lower frequency. Indeed some analyses of source counts with shallower LOFAR surveys in the LH and Boo fields claimed in the past a better consistency with a $k=4$ \citet{windhorst90} median size -- flux scaling relation (\citealt{mahony16};  \citealt{retanamontenegro18}). It is interesting to note, however, that the LoTSS final catalogues are characterized by smaller median sizes than the raw catalogues at their faint end ($S_{\rm{1.4GHz}}\la 5$ mJy), indicating that confusion significantly affects the measured sizes of the faintest sources, and that a significant number of faint sources were deblended. On the other hand, the final catalogues tend to be characterised by larger sizes at the bright end ($S_{\rm{1.4GHz}}\ga 100$ mJy), likely as a consequence of the association of multiple components into single sources, after visual inspection of the radio/optical images (see Sect~\ref{sec-debl}). After accounting for these effects, LoTSS median sizes (golden filled circles) are consistent with the \citet{windhorst90} $k=2$ size -- flux relation up to $S_{\rm{1.4GHz}}\sim 2$ mJy. Then they smoothly increase and become consistent with the \citet{bonaldi19} relation at $S_{\rm{1.4GHz}}\ga 10$ mJy. At $S_{\rm{1.4GHz}}\ga 100$ mJy the LoTSS source median sizes show large uncertainties. At these large flux densities also the \citet{bonaldi19} relation is poorly determined, being based on a simulated catalogue covering a similar area to the one covered by the LOFAR deep fields (25 deg$^2$). It is interesting to note, however, that both are consistent with the \citet{windhorst90} relation. Based on all the above considerations, a good description of the observed median sizes can be obtained by the following analytical form, which assumes the \citet{windhorst90} relation with a varying $k=k(S)$, i.e.:
\begin{align}
k= 
\begin{cases}
3.5 - 1.5\times \rm{exp}(-S_{\rm{1.4GHz}}/2) &  S_{\rm{1.4GHz}}< 4.5\\
2 + 1.5\times \rm{exp}(-S_{\rm{1.4GHz}}/200) & S_{\rm{1.4GHz}}\geq 4.5
\end{cases}
\label{eq:w90ks}
\end{align}
where $S_{\rm{1.4GHz}}$ is expressed in mJy (see solid line in Fig.~\ref{fig:median-size}). 

\begin{figure*}[!t]
\begin{center}
\vspace{-1.5cm}
\resizebox{0.9\hsize}{!}{
\includegraphics[angle=0]{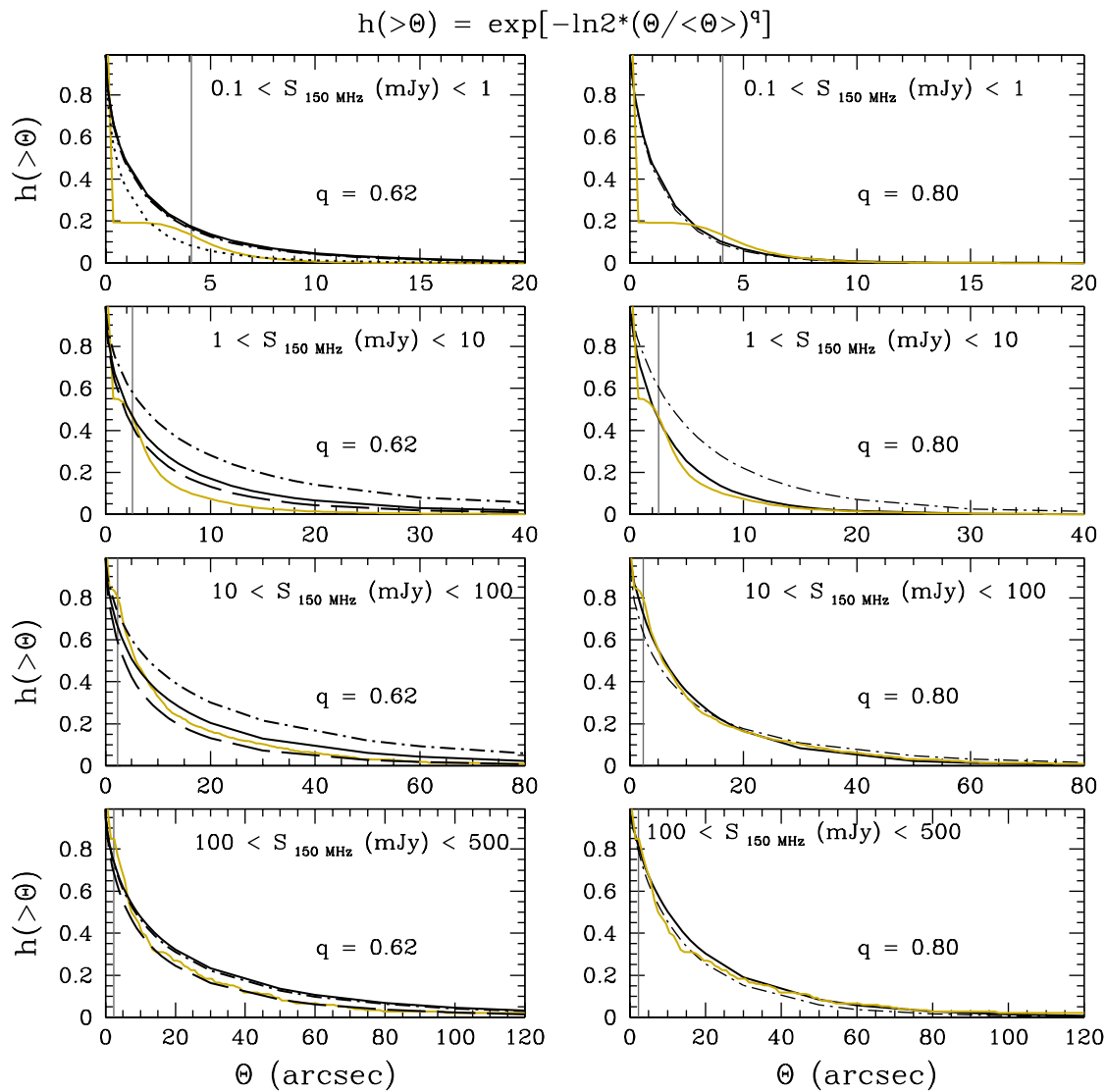}
}
\vspace{-4cm}
\caption[]{Source size cumulative distribution of final catalogues (yellow solid line) in four 150 MHz flux density bins (\textit{top} to \textit{bottom}).  The vertical grey lines in all panels provides an approximate indication of the minimum intrinsic angular size to which the observed distributions can be considered reliable (most of the sources below this line cannot be reliably deconvolved and they are conventionally assigned $\Theta=0$). Also shown for comparison are various realizations of  the cumulative distribution function described by Eq. \ref{eq:cumdistr}. The two columns correspond to two different values for the function exponent $q$: the original one proposed by \citet{windhorst90} ($q=0.62$) on the \textit{left}, and a steeper one ($q=0.80$) on the \textit{right}. In addition we also vary the  median size -- flux relation. In particular we assume the original \citet{windhorst90} relation (black long-dashed line), the revised versions  with flux-dependent $m$ and $k$ parameters, as described by Eqs.~\ref{eq:revw90} and~\ref{eq:w90ks} (black dotted and solid lines
respectively) and the one describing the T-RECS catalogues (\citealt{bonaldi19}; black dot-dashed line). All such realizations are shown on the left; on the right we only show the realizations obtained using the \citet{bonaldi19} and the revised  \citet{windhorst90} $k=k(S)$ relations. 
\label{fig:cumsize}}
\end{center}
\end{figure*}

Another important consistency check regards the angular size distribution of the sources. 
Figure~\ref{fig:cumsize} shows the cumulative size distributions of the final catalogues combined together, in four flux density bins (yellow solid lines). Such distributions can be considered reliable only down to a flux-dependent minimum intrinsic size (see vertical grey lines), below which most of the sources  cannot be reliably deconvolved and they are conventionally assigned $\Theta=0$. The observed distributions are compared with various realizations of the cumulative distribution function described by Eq. \ref{eq:cumdistr}, obtained by varying either the function exponent $q$ (left and right columns respectively) or the assumed median size -- flux relations (see various black lines). The original function proposed by \citet{windhorst90} (Eq. \ref{eq:cumdistr} with $q=0.62$, see left column)  does provide a good approximation of the observed distributions,  when assuming the original $\Theta_{\rm med} - S$ relation  described by Eq. \ref{eq-sizew90}, only at flux densities $S_{\rm 150 MHz}\ga 10$ mJy (see long-dashed lines). This is perhaps not surprising considering that this relation was calibrated at 1.4 GHz down to a few mJy fluxes. At the lowest flux densities ($S_{\rm 150 MHz}\la 1$ mJy) we need to assume a steepening of the parameter $m$ (see Eq. \ref{eq:revw90}), to get a good match with observations  (dotted line in the top left panel). This is consistent with what proposed for higher frequency  deep surveys (as discussed earlier in this Section). At intermediate fluxes ($S_{\rm 150 MHz}\sim  1-10$) mJy, on the other hand, none of the discussed median size -- flux relations can reproduce the observed  size distribution (see second-row panel on the left). 
It is interesting to note, however, that if we assume a steeper exponent for the distribution function described by Eq. \ref{eq-sizew90} (i.e. $q=0.80$), we get a very good match with observations at all fluxes, when assuming a flux-dependent scaling factor ($k=k(S)$; see Eq. \ref{eq:w90ks}) for the \citet{windhorst90} median size -- flux relation (black solid lines on the right). 
The median sizes derived from the 
T-RECS simulated catalogues (\citealt{bonaldi19}) also provide good results for $q=0.80$ (dot-dashed lines on the right), 
except again at intermediate fluxes 
($S_{\rm 150 MHz}\sim 1-10$), where they show strong discrepancies with observations also in 
Fig. \ref{fig:median-size}. This seems to indicate that the number density of extended radio galaxies in this flux density range is  over-estimated in the T-RECS simulated catalogues. 

\subsection*{Correction for Resolution Bias}\label{sec:corrresbias}
 The correction factor \textit{c} that needs to be applied to the source counts to account for the resolution bias can be defined as (\citealt{prandoni01}):
\begin{equation}
\textit{c}= 1/[1-\textit{h}(>\Theta_{\rm{lim}})]
\end{equation}
where $\textit{h}(>\Theta_{\rm{lim}})$ takes the form of the  integral of the angular size distribution proposed by \citet[see Eq. \ref{eq:cumdistr}]{windhorst90}, and  $\Theta_{\rm{lim}}$ is the limiting angular size above which the catalogues are expected to be incomplete. Following \citet{prandoni01}, this is defined as:
\begin{equation}
\Theta_{\rm{lim}}=\textit{\rm{max}}[\Theta_{\rm{min}},\Theta_{\rm{max}}]
\end{equation}
where $\Theta_{\rm{max}}$ and $\Theta_{\rm{min}}$ are as defined  in Eqs.~\ref{eq:thetamax} and~\ref{eq:thetamin} respectively. We notice that $\Theta_{\rm{lim}}$ is always equal to $\Theta_{\rm{max}}$, except for the lowest flux bins, where $\Theta_{\rm{max}}$ becomes unphysical (i.e. tends to zero). $\Theta_{\rm{min}}$ accounts for the effect of having a finite restoring beam size (that is $\Theta_{\rm{lim}}>0$ at the survey limit) and a deconvolution efficiency which varies with the source peak flux density (see \citealt{prandoni01} for more details). 

\begin{figure*}[!th]
\begin{center}
\vspace{-1cm}
\resizebox{0.45\hsize}{!}{
\includegraphics[angle=0]{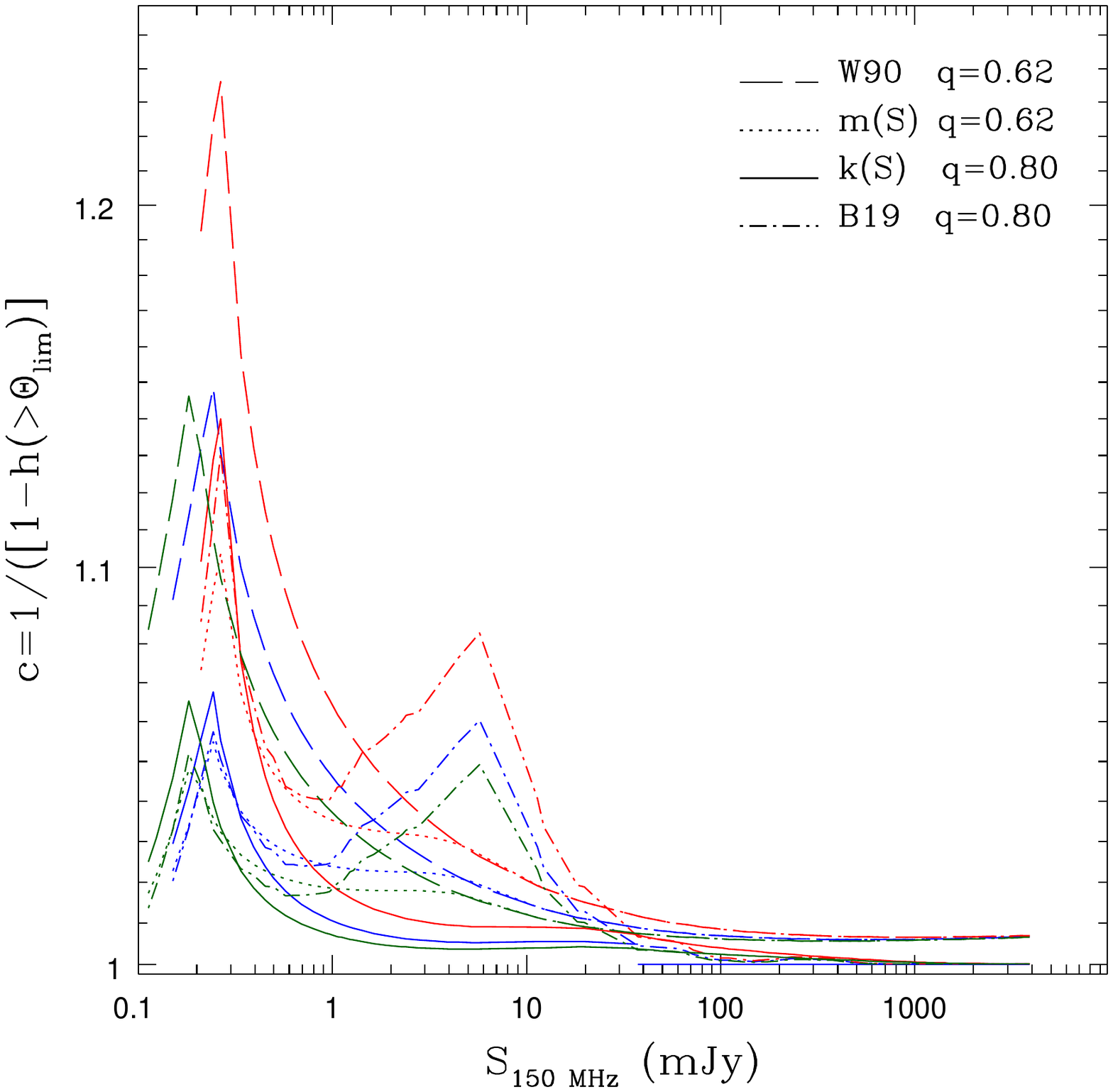}
}
\resizebox{0.45\hsize}{!}{
\includegraphics[angle=0]{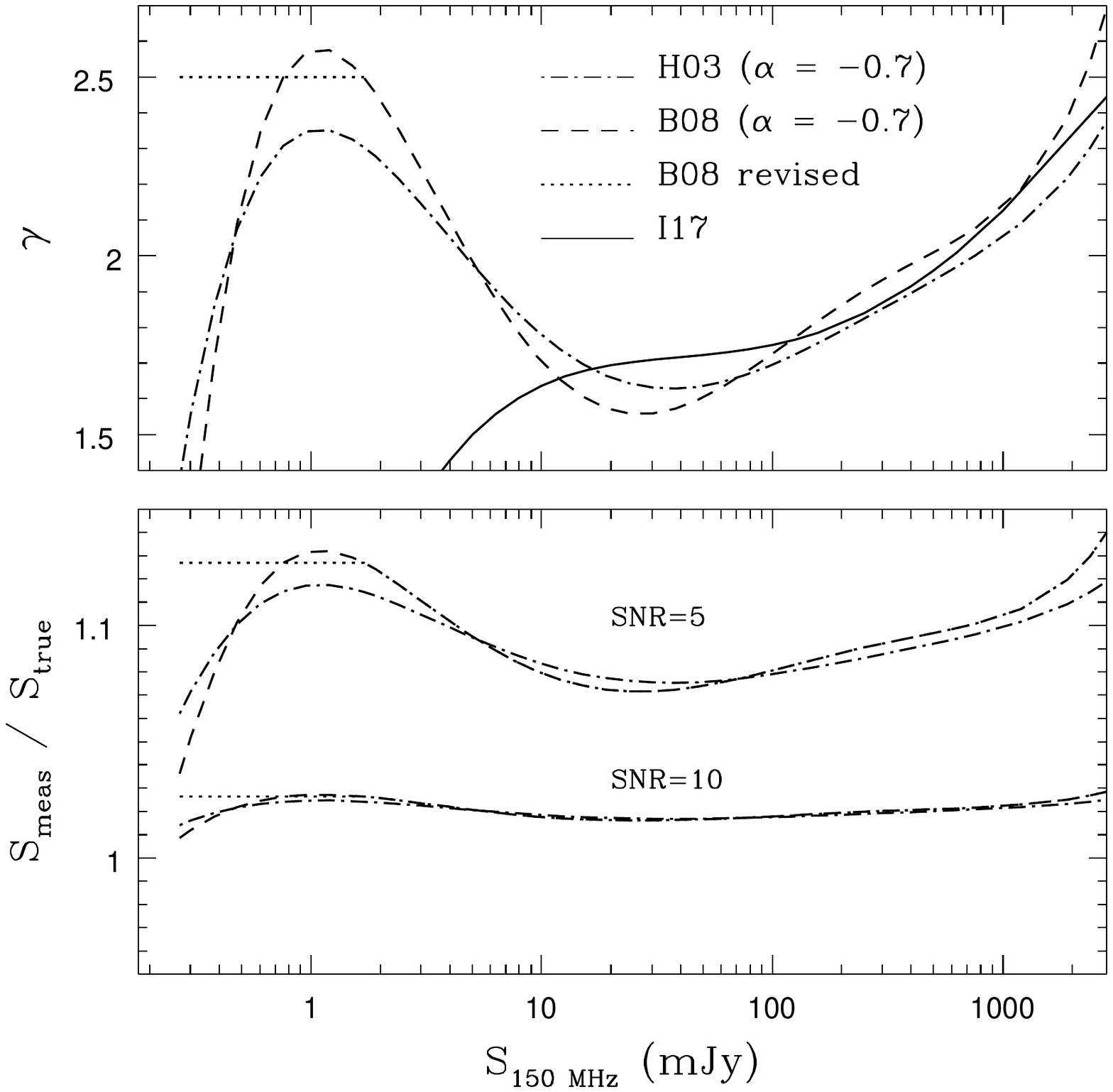}
}
\vspace{-2cm}
\caption{{\it Left:} Flux-dependent correction to be applied to source counts to account for incompleteness due to resolution bias, for four median size - flux relations: the one derived from the simulated T-RECS catalogues (\citealt{bonaldi19}; dot-dashed lines), the one proposed by \citet[][long-dashed lines)]{windhorst90}, the revised version with $m=m(S)$, which better describe source sizes at 1.4 GHz sub-mJy fluxes (see Eq.~\ref{eq:revw90}, dotted lines), and the revised version with $k=k(S)$ proposed by us (Eq.~\ref{eq:w90ks}, solid lines). We also vary the $q$ exponent of the integral distribution function. Based on our analysis of the source size distribution (see Fig. \ref{fig:cumsize} and related discussion), we assume the  original value  proposed by \citet[$q=0.62$]{windhorst90} for the \citet{windhorst90} median size -- flux relation and for  the revised version with $m=m(S)$. We assume a steeper $q=0.80$ for the revised version with $k=k(S)$ and for the \citet{bonaldi19} relation (see  legenda).
Different colors refer to different fields: LH (blue), Boo (red), EN1 (green). The corrections account for noise variations in the masked images through an empirical  relation between source flux and source signal-to-noise ratio, calibrated for each field (we assume here the median noise of the masked images; see last column of Tab. \ref{tab:image-properties}).  {\it Right:}  Eddington bias for different underlying number-count distributions, as illustrated in the top panel: source counts' slope ($\gamma$; $dN/dS \sim S^{-\gamma}$) derived from the sixth-order polynomial fit proposed at 1.4 GHz by a) \citet{hopkins03} (dot-dashed line) and b) \citet{bondi08} (dashed line), both converted to 150 MHz assuming a spectral index $\alpha= -0.7$; we also show a revised version of the \citet{bondi08} fit, which assumes a constant Euclidean slope ($\gamma=2.5$) from 2 mJy all the way down to 0.1 mJy (dotted line). The polynomial fit proposed by \citet{intema17} at 150 MHz and valid only for the bright end of the counts is also shown for reference (solid line). The flux boosting  ($S_{\rm meas}/S_{\rm true}$) corresponding to the three cases illustrated above is shown in the bottom panel for two different source signal-to-noise ratios: SNR=5 and SNR=10. 
\label{fig:bias}}
\end{center}
\end{figure*}

Figure~\ref{fig:bias} (left panel) shows the correction factor derived assuming the  median size -- flux relations discussed above, combined with appropriate values of the $q$ exponent in Eq. \ref{eq:cumdistr}, based on our analysis of the source size distribution (see Fig. \ref{fig:cumsize} and related discussion). A caveat to keep in mind is that the resolution bias correction does depend on both the source flux density and the noise value at the source position (and/or the source signal-to-noise ratio; see Eqs.~\ref{eq:thetamax} and~\ref{eq:thetamin}). The corrections presented  in Fig.~\ref{fig:bias} (left panel) account for local and radial variations of the noise through empirical relations between source flux and local noise or signal-to-noise ratio, specifically derived for each field. Such relations describe average trends only, and hence the corrections presented here should be considered as indicative. The corrections effectively applied to the counts are based on the actual source flux density, noise and signal-to-noise ratio distributions. It is interesting to note, however, that, as a consequence of  radially-increasing noise (and/or limited dynamic range around bright radio sources), the correction factor $c$ does not necessarily converge to 1 at large flux densities. As shown in Fig.~\ref{fig:bias}, in the masked regions of our fields this only happens when assuming the shallower integral distribution function ($q=0.62)$. For the steeper one ($q=0.80$), the expected number density of very extended sources is  small, and resolution bias effects become negligible at $S_{\rm 150 MHz} \ga 500$ mJy.

\section{Eddington bias}\label{sec:eddbias}
While correcting for resolution bias is important to account for missed resolved sources, Eddington bias (\citealt{eddington13,eddington40}) should be taken into account to get an unbiased census of unresolved sources. Due to random measurement errors the measured peak flux densities will be  redistributed around their true value. In presence of a source population  which follows a non-uniform flux distribution, this will result in a redistribution of sources between number-count flux density bins. The way the sources are redistributed depends on the slope of source counts. If the source number density increases with decreasing flux, the flux densities tend to be boosted and the probability to detect a source below the detection threshold is higher than the probability to miss a source above the threshold, artificially boosting the detection fraction. As a consequence the catalogue incompleteness at the detection threshold is also biased.

There are two main approaches to correct for Eddington bias, both requiring an assumption about the true underlying source counts distribution (see \citealt{hales14a} for a full discussion): one can build the source counts using the boosted flux densities and then apply a correction to each flux density bin, or one can correct the source flux densities, before deriving the counts.
As demonstrated by \citet{hales14b} the two approaches give very similar and consistent results, and we decided to follow the latter approach. A maximum likelihood solution for the true source flux density can be defined as follows (see \citealt{hales14a} and references therein):
\begin{equation}
    S_{\rm true} = \frac{S_{\rm meas}}{2} \cdot \left(1+ \sqrt{1-\frac{4\gamma}{\rm SNR^2}}\right)
\end{equation}
where $\gamma=\gamma(S)$ is the slope of the counts at the given flux density ($dN/dS \sim S^{-\gamma}$), and SNR is the source signal-to-noise ratio. The slope of the counts can be modeled from empirical polynomial fits of the observed counts:
\begin{equation}
    \log\left(\frac{dN(S)}{dS} S^{2.5}\right) = \sum_{i=0}^n a_i \left(\log S \right)^i \; .
\label{eqn:polyfit}
\end{equation}
It is then easy to demonstrate that:
\begin{equation}
    \gamma = 2.5- \sum_{i=0}^n i\cdot a_i \left(\log S \right)^{i-1}  \; .
\end{equation}
In order to derive $\gamma$ we can use one of the several counts' fits available in the literature. \citet{intema17} derived a fifth-order polynomial fit which describes the 150 MHz normalized counts of the TIFR GMRT Sky Survey (TGSS-ADR1), but this fit is only valid down to a flux limit of 5 mJy. The deepest fits available in the literature have been obtained at 1.4 GHz. We start by exploring the sixth-order ($n=6$) polynomial fits obtained by \citet{hopkins03} and \citet{bondi08} for 1.4 GHz normalized source counts (converted to 150 MHz using $\alpha=-0.7$\footnote{We note that assuming a single spectral index is a crude approximation. In principle we should account for the intrinsic scatter  in the spectral index distribution of the sources, as well as for possible deviations of the mean spectral index with flux density, due to the varying relative contribution of the individual source populations. 
Such an approximation is however acceptable, since the largest uncertainties in this analysis come from the assumptions on the counts' slope at the faintest fluxes, which is very poorly known.}).  
The two cases are illustrated in the top right panel of Fig.~\ref{fig:bias} (dot-dashed and dashed lines respectively), where the derived counts' slope is shown. Both cases are consistent with the \citeauthor{intema17} 150 MHz fit (indicated by the solid line) at bright flux densities ($S>100$ mJy), while significant discrepancies are observed at fainter fluxes, where the deeper 1.4 GHz fits better describe the well-known flattening of the normalized counts. 
Both the 1.4 GHz fits show an increasing slope below 10 mJy, reaching a maximum around 1 mJy. This maximum is more pronounced  in the case of \citet{bondi08}, and is consistent with an Euclidean slope of $\gamma \sim 2.5$. At $S<1$ mJy both slopes show a rapid drop. The reality and strength of this drop is unclear, as this is the flux regime where the fits are less reliably constrained. We then explore a third case, i.e. a modification of the \citeauthor{bondi08} fit, which assumes a constant Euclidean slope at flux densities $S<2$ mJy. This 
represents an extreme scenario, which however might be favoured by the recent 150 MHz source counts modeling proposed by \citet{bonaldi19}, that indicates a flatter slope in the flux range $0.1-1$ mJy. This last case is illustrated by the dotted line  in Fig.~\ref{fig:bias} (top right panel).  The flux density boosting expected for the three aforementioned scenarios is illustrated in the bottom right  panel of Fig.~\ref{fig:bias} for two signal-to-noise ratio values, SNR=5 and SNR=10. 

Once the point source flux densities are corrected for Eddington bias, we can obtain an estimate of the catalogue incompleteness at the detection threshold through the use of Gaussian Error Functions (ERF). 


\begin{figure*}[!htb]
\vspace{-1cm}
\begin{center}
\resizebox{0.78\hsize}{!}{
\includegraphics[angle=0,clip]{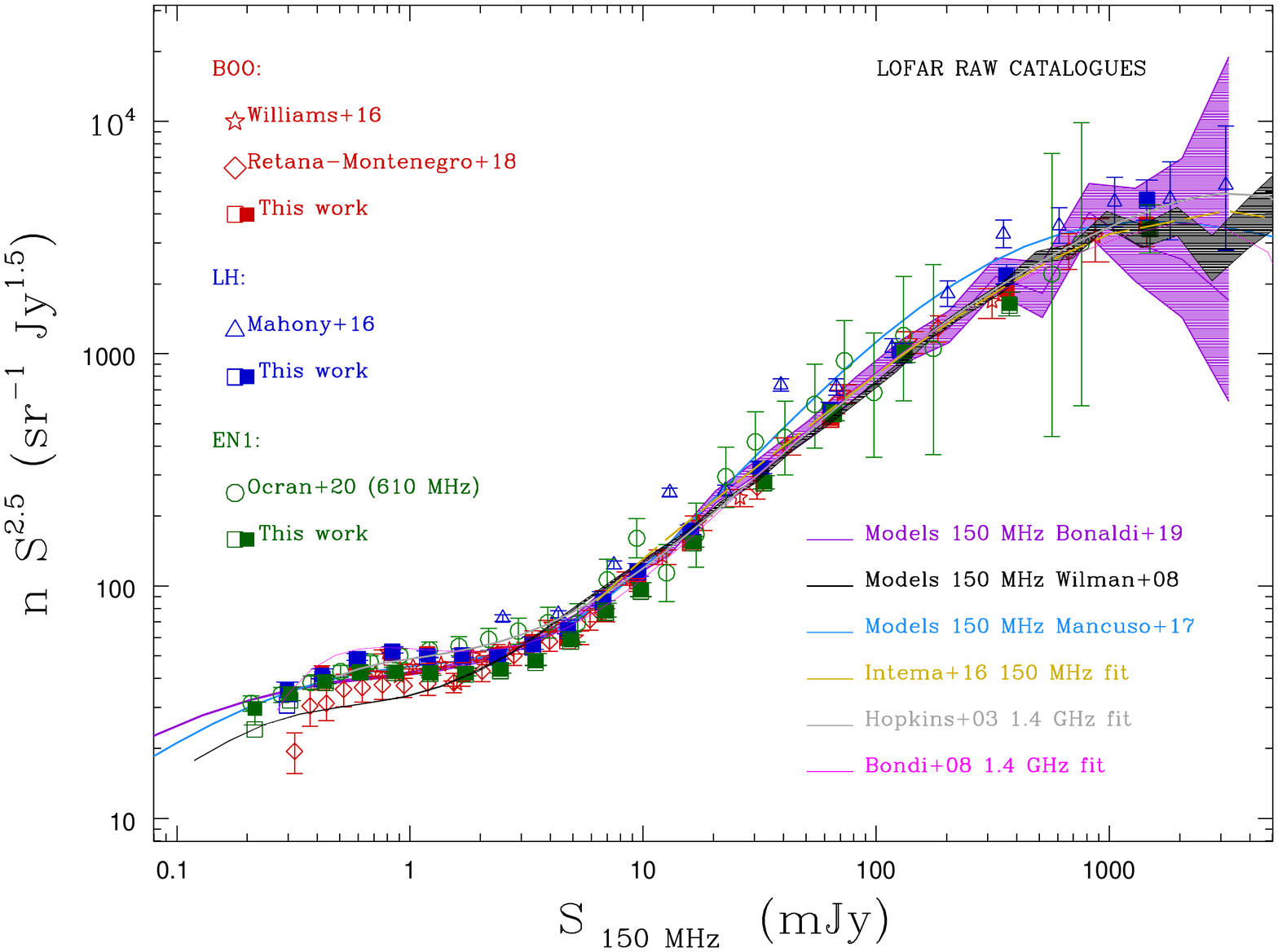}
}
\vspace{-1cm}

\resizebox{0.78\hsize}{!}{
\includegraphics[angle=0,clip]{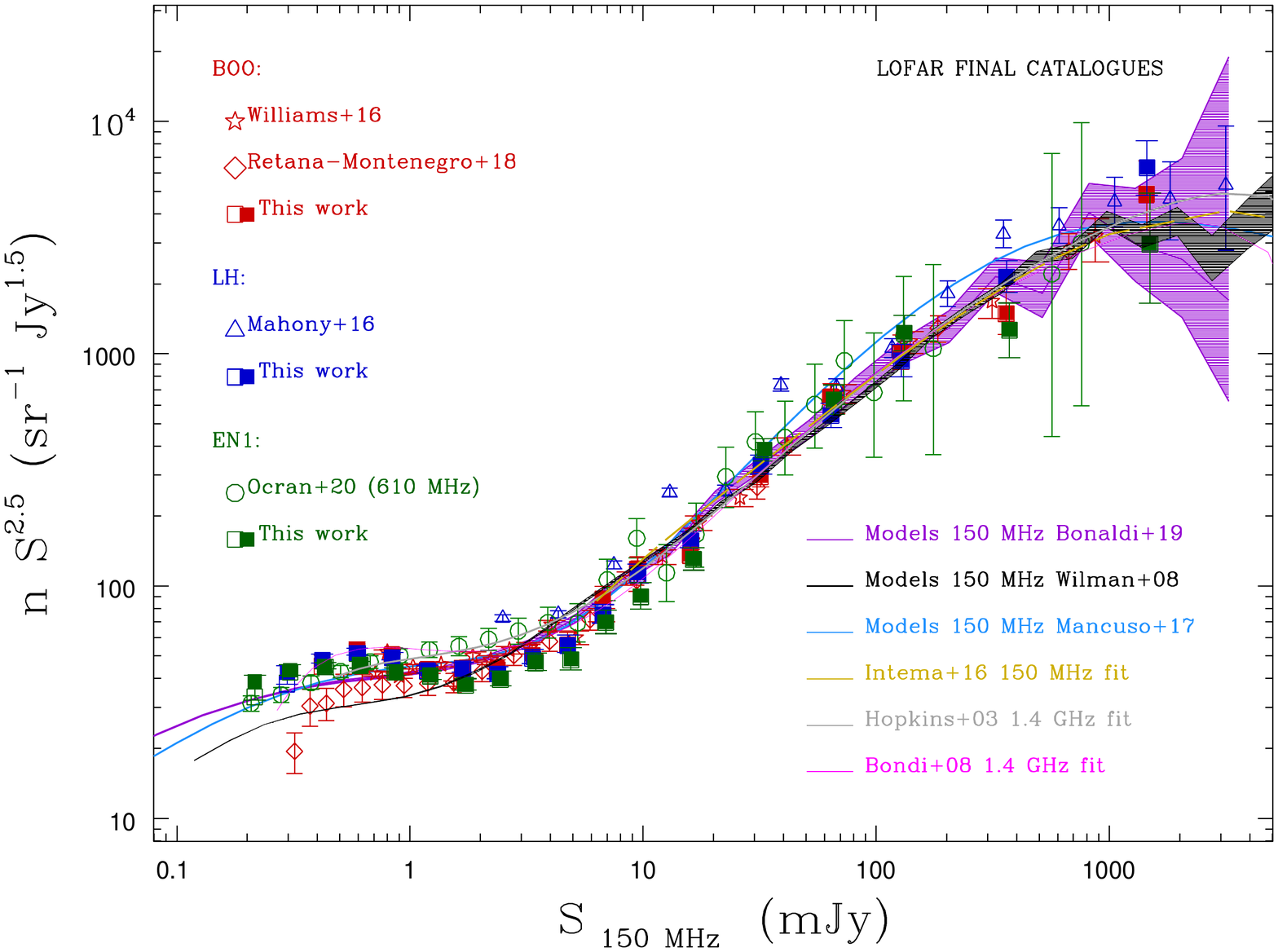}
}
\caption[]{Normalized 150 MHz differential source counts in the three LoTSS Deep Fields, as derived from raw ({\it top}) and final ({\it bottom}) catalogues (filled squares). Error bars correspond to the quadratic sum of Poisson and systematic errors. Also shown are the counts obtained without applying the corrections discussed in Sects. \ref{sec:resolution-bias} and \ref{sec:eddbias} (empty squares). The counts are derived by using total flux densities for both point and extended sources.  In both figures, \cite{wilman08}, \cite{bonaldi19} and \citet{mancuso17} 150 MHz models are shown for comparison, as well as other existing 150~MHz counts' determinations in the same fields (see legend). Since published  150 MHz counts are missing for EN1, we show a recent determination obtained at 610 MHz (\citealt{ocran20}) and rescaled to 150 MHz, assuming $\alpha=-0.7$. Also shown are the counts' best fits discussed in Sect. \ref{sec:eddbias}.}
\label{fig:source_count_raw_debl}
\end{center}
\end{figure*}

\begin{figure*}
\vspace{-0.5cm}
\resizebox{0.7\hsize}{!}{\includegraphics[width=12cm,clip]{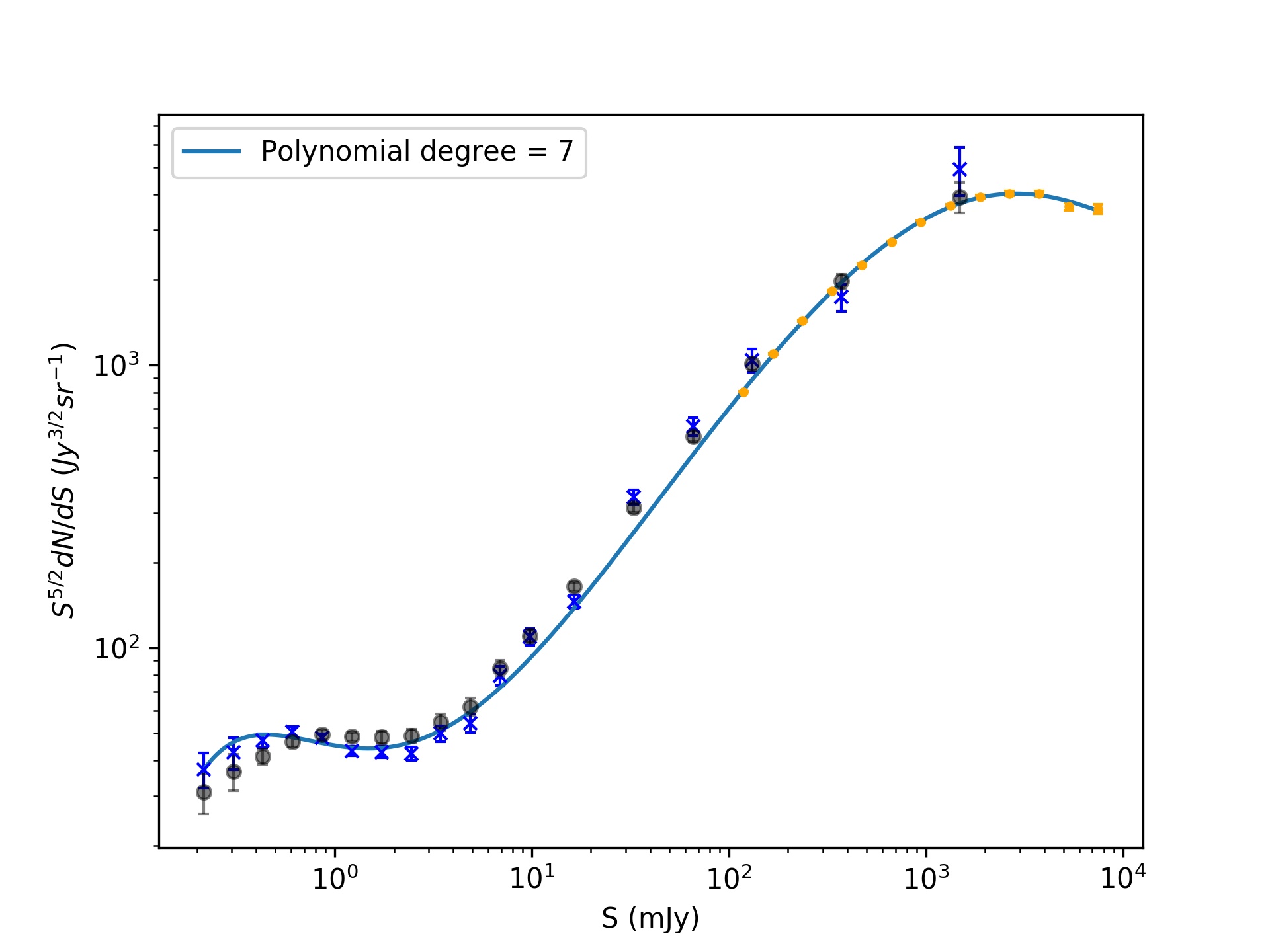}}
\hfill
\begin{minipage}[t]{55mm}
\vspace{-6cm}
\caption{150~MHz Euclidean normalized differential source counts as derived from the LoTSS Deep Fields: raw catalogue are indicated by transparent black circles and final catalogue by blue crosses). Also shown are the counts obtained from the TGSS-ADR1 (\citealt[orange filled circles]{intema17}), which better describe the counts' bright end. Over-plotted is the best fit obtained by modeling the counts in the log-log space with a $7$-th order polynomial function, according to Eq. \ref{eqn:polyfit} (see Table \ref{tab:polyfit_values} for the values of the best-fit coefficients and associated errors). 
}
\label{fig:polfit}
\end{minipage}
\end{figure*}

\begin{table}[tb]
    \caption{150~MHz normalized differential radio-source counts as derived from combining the raw and final catalogues of the three LoTSS Deep Fields.  }
   \label{tab:comb_counts}
    \centering
   \begin{tabular}{|c|c|c|}
   \hline
        $<S>$ & $N(\rm{raw})^{+\sigma_{\rm tot}}_{-\sigma_{\rm tot}}$  & $N(\rm {final})^{+\sigma_{\rm tot}}_{-\sigma_{\rm tot}}$ \\
         \hline
      0.22 & $30.92_{-6.65}^{+2.20}$  & $37.15_{-7.92}^{+2.59}$   \\
      0.31 & $36.57_{-7.04}^{+3.13}$  & $42.65_{-7.57}^{+3.40}$   \\
      0.43 & $41.29_{-2.17}^{+3.13}$  & $47.08_{-1.86}^{+3.40}$  \\
      0.61 & $46.42_{-0.52}^{+3.16}$  & $50.55_{-0.62}^{+3.40}$    \\
      0.86 & $49.39_{-0.42}^{+3.15}$   & $47.97_{-0.68}^{+2.85}$    \\
      1.22 & $48.48_{-0.53}^{+3.04}$  & $43.17_{-0.83}^{+2.23}$  \\
      1.73 & $48.24_{-0.67}^{+4.13}$   & $42.77_{-1.07}^{+2.77}$  \\
      2.45 & $48.73_{-0.87}^{+4.76}$  & $42.30_{-1.37}^{+3.30}$  \\
      3.46 & $54.60_{-1.19}^{+6.20}$   & $49.81_{-1.93}^{+4.45}$  \\
      4.89 & $61.95_{-1.64}^{+7.08}$  & $54.21_{-2.61}^{+5.71}$  \\
    6.92 &  $84.37_{-2.47}^{+9.42}$  & $79.79_{-4.10}^{+8.09}$ \\
    9.79 &  $110.2_{-3.7}^{+8.2}$  & $109.6_{-6.2}^{+8.8}$ \\
    16.5 & $164.5_{-4.6}^{+6.5}$  & $145.9_{-7.5}^{+8.3}$ \\
    32.9 & $312.8_{-10.7}^{+11.2}$ & $341.5_{-19.2}^{+20.3}$  \\
     65.8 & $560.6_{-24.1}^{+25.2}$ & $607.1_{-42.9}^{+46.1}$   \\
      132 & $1012_{-55}^{+58}$ & $1041_{-95}^{+103}$    \\
      372 & $1977_{-115}^{+123}$  & $1739_{-183}^{+203}$  \\
      1489 & $3932_{-458}^{+515}$  & $4921_{-870}^{+1027}$  \\
      \hline
    \end{tabular}
    \tablefoot{$<S>$ is the geometric mean of the respective flux density bin, expressed in mJy;  $N$(raw) and $N$(final) indicates the normalized source counts obtained from the raw and final catalogues respectively (in $Jy^{1.5}sr^{-1}$); $\pm \sigma_{\rm tot}$ are the total errors on the counts, estimated as the quadratic sum of Poissonian and systematic errors. We note that only EN1 and LH sources contribute to the first flux density bin. }
    \\
\end{table}

\section{Differential source counts}\label{sec:counts}

\begin{figure*}[tb]
\begin{center}
\vspace{-1cm}
\resizebox{0.78\hsize}{!}{
\includegraphics[angle=0,clip]{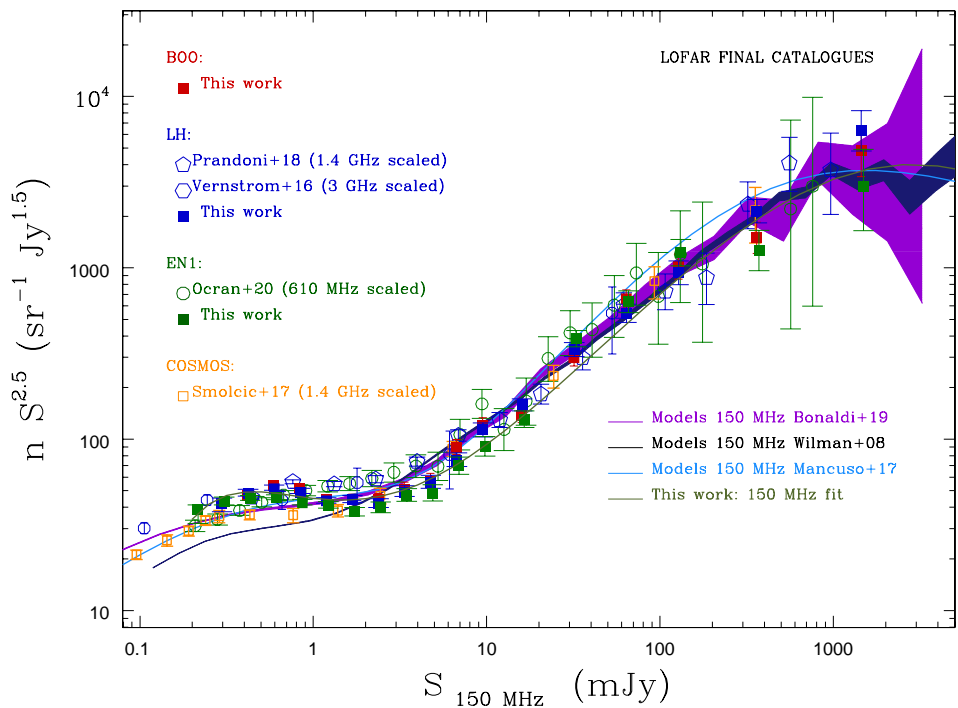}
}
\caption[]{Normalized 150 MHz differential source counts in the three LoTSS Deep Fields, as derived from final catalogues (filled squares), together with their best fit (grey solid line). They are compared to counts derived from higher frequencies surveys (see legend), and rescaled to 150 MHz by assuming $\alpha=-0.7$. Error bars correspond to the quadratic sum of Poisson and systematic errors.  Also shown are the \cite{wilman08}, \cite{bonaldi19} and \citet{mancuso17} 150 MHz models. The counts are derived by using total flux densities for both point and extended sources.
\label{fig:source_count_all}}
\end{center}
\end{figure*}

The differential source counts, normalised to a non-evolving Euclidean model, obtained from the LoTSS Deep Fields are shown in Figure \ref{fig:source_count_raw_debl}, together with other count determinations obtained in the same fields from previous low-frequency surveys (see legend). The top and bottom panels refer to counts derived from the raw and final catalogues, respectively (see filled boxes). The source counts obtained from the final catalogues are reported in tabular form in the Appendix (Tables~\ref{tab:counts_LH}, \ref{tab:counts_Boo} and \ref{tab:counts_EN1}, for LH, Boo and EN1 fields respectively), and are also shown in Fig. \ref{fig:source_count_all}, where they are compared to counts extrapolated from higher frequencies. In deriving the counts we applied a `fiducial' model for the systematic corrections described in Sects.~\ref{sec:resolution-bias} and~\ref{sec:eddbias}. Specifically we assumed the \citet{windhorst90} size -- flux relation with $k=k(S)$ (Eq. \ref{eq:w90ks}), in combination with a `steep' ($q=0.80$ in Eq. \ref{eq:cumdistr}) integral size distribution, to estimate the resolution bias, and we assumed the \citet{bondi08} source counts best fit to estimate the Eddington bias. The uncertainties associated with such assumptions are factored into  systematic error terms (see Sys$^-$ and Sys$^+$ columns in the counts' tables), that are defined as the maximum discrepancy between the `fiducial' counts and those obtained assuming the other discussed models (shown in Fig. \ref{fig:bias}). Also shown in Fig. \ref{fig:source_count_raw_debl} are the counts obtained from the three LoTSS fields before applying the corrections for resolution and Eddington bias (empty boxes). As expected such corrections are only relevant for the lowest flux density bins. 
We cut the source counts at a threshold of $\sim 7\sigma_{\rm med}$, where systematic errors dominate over Poissonian (calculated following \citealt{gehrels86})  by factors $\sim 5-10$.  

The normalized 150 MHz source counts derived from the three LoTSS Deep Fields are in broad agreement with each other, and show the well known flattening at $S\la$ few mJy. The observed field-to-field variations for the counts derived from the final catalogues are typically of the order of a few percent at sub-mJy fluxes, and of 5-10\% at flux densities $1-10$ mJy, with EN1 showing somewhat larger deviations (up to 15-20\% in some flux density bins). This is consistent with expectations from sample variance for surveys covering areas of $5-10$ deg$^2$ (\citealt{heywood13,prandoni18}). We notice that all of the other 150 MHz counts shown in Fig. \ref{fig:source_count_raw_debl} come from previous shallower LOFAR observations, but the catalogues did not go through the source  association and deblending post-processing described in Sect. \ref{sec-debl}. This explains why determination of previous source counts are more in line with those we obtained from raw catalogues (see top panel of Fig. \ref{fig:source_count_raw_debl}). The largest discrepancies are observed in the Boo field with the counts by \citet{retanamontenegro18}. The systematically higher counts that we obtain at sub-mJy flux densities may be  the result of
the new calibration and imaging pipeline, that produces higher fidelity, higher dynamic range images (paper I). Another source of systematic differences may be a residual offset in the absolute flux scale. 

When comparing the source counts derived from raw and final catalogues (top and bottom panels of Fig. \ref{fig:source_count_raw_debl}), we notice a very interesting feature: the latter show a much more pronounced drop at flux densities around a few mJy, which results in a more prominent `bump' in the sub-mJy regime.  
For a more quantitative analysis of this feature we have combined the sources in the three fields and produced a combined version of the source counts. This allows us to  smooth out the aforementioned field-to-field variations, as well as reduce the scatter at bright flux densities, where Poissonian errors dominate. The combined counts derived from raw and final catalogues are shown in Fig. \ref{fig:polfit} and listed in Tab.\ref{tab:comb_counts}. We notice that in this case we included LH and Boo sources down to a $5\sigma$ flux density limit, to increase the statistics available in the first two flux density bins. 
From the comparison of the raw and final counts we see that the latter are systematically lower by a factor  $5-15\%$  in the range $S\sim 1-10$ mJy. This deficiency appears to be  counterbalanced by an $8-16\%$ excess at flux densities $0.2-0.6$ mJy.
This form of compensation is likely the result of source deblending (see Sect. \ref{sec-debl} for more details). Indeed, for each field we have $\sim 1000s$ deblended sources (see paper III), and deblending is mostly effective at the faint end of the counts, where it contributes to the increase of the number of the sources, through the splitting of confused (brighter) sources into several fainter ones. Artifact removal, on the other hand, has probably a limited effect.  Artifacts mostly affect the flux density range $2-20$ mJy\footnote{This was verified by running PyBDSF on the negative raw map (i.e. the raw map multiplied by -1), with the same input parameters used to extract the source catalogue (see e.g. \citealt{prandoni18}).}, but only $<100$ sources per field are confirmed as such (see paper III). The fact that source cataloguing processes can significantly affect the derived number counts is well known. Significant discrepancies are observed, for instance, when comparing counts derived from source catalogues with those directly obtained from Gaussian components (see e.g. \citealt{hopkins03}). While PyBDSF creates a source catalogue, our  association and deblending procedure aims at further combining Gaussian components together (association) or splitting  them into several independent sources (deblending), hence modifying the source statistics and flux distribution of the original PyBDSF catalogue.

As a final remark we note that none of the best fits from previous source counts (shown in Fig. \ref{fig:source_count_raw_debl})  provide a good description of the pronounced `drop and bump' feature that we observe at 150 MHz. We have therefore derived a new  fit that  better matches the faint end of the 150 MHz counts derived from our final catalogues. The slope of the counts is modeled by a $7$-th order polynomial function defined in the log-log space, according to Eq. \ref{eqn:polyfit} (see Sect. \ref{sec:eddbias} for more details). To better constrain the bright end of the counts, where the LoTSS Deep Fields provide poor statistics, we have included the counts derived from the TGSS-ADR1 (\citealt{intema17}), up to a flux density of 10 Jy (beyond that the TGSS-ADR1 itself is affected by poor statistics). The resulting coefficient values and their uncertainties are listed in Table \ref{tab:polyfit_values}; the fitted curve is shown in Fig. \ref{fig:polfit}.

\begin{table}[tb]
    \caption{Coefficients for $7$-th order polynomial function defined by Eq. \ref{eqn:polyfit}, which best-fit the LoTSS and TGSS-ADR1 150 MHz normalized source counts in the flux density range: $0.2$ mJy $- 10$ Jy. The polynomial fit is shown in Figure \ref{fig:polfit}.}
   \label{tab:polyfit_values}
    \centering
   \begin{tabular}{|c|c|c|}
   \hline
      Coefficient & Value & Error ($\pm$) \\
         \hline
      $a_0$ &  1.655 & 0.017 \\ 
      $a_1$ & -0.1150  & 0.0654 \\
      $a_2$ &  0.2272 & 0.097 \\
      $a_3$ &  0.51788 & 0.18616 \\
      $a_4$ & -0.449661 & 0.185032 \\
      $a_5$ &  0.160265 & 0.084983 \\
      $a_6$ & -0.028541 & 0.018280 \\
      $a_7$ & 0.002041 & 0.001497 \\ 
          \hline
    \end{tabular}
\end{table}

\subsection*{Qualitative comparison with higher frequency counts }

Our source counts are the deepest available at 150 MHz. Their faint end can therefore only be probed against counts obtained from higher frequency surveys of similar depth. In Fig. \ref{fig:source_count_all} we show the normalized differential counts in the three LoTSS Deep Fields, as derived from final catalogues, together with counts derived from higher frequencies surveys over the same regions of the sky, and rescaled to 150 MHz by assuming $\alpha=-0.7$.
We also add the COSMOS 3GHz Large project (\citealt{smolcic17}), which provides the deepest counts to date, over a degree-scale field. None of the higher frequency counts show the pronounced `drop' at mJy flux densities, that we observe at 150 MHz. The only exception is COSMOS, which is in agreement with our 150 MHz counts at $S\ga 1$ mJy, but falls below them at lower flux densities. We note that sample variance can in principle explain variations $\ga 15-25\%$ in the range $1-10$ mJy for surveys covering areas $\la 2$ deg$^2$ (\citealt{heywood13}), like COSMOS and the ones by \citet{ocran20} and \citet{vernstrom16} in EN1 and LH respectively, but can hardly justify the observed discrepancies with \citet{prandoni18} sample, which covers more than half ($\sim 6.6$ deg$^2$) of the LoTSS LH field. In addition, sample variance is expected to be of the order 5-10\% at sub-mJy fluxes for a sample like COSMOS, i.e. smaller than what observed. It is then likely that survey systematics additionally contribute to the observed field-to-field variations. In this respect we note that the excess at mJy flux densities is significantly mitigated when comparing the higher frequency counts to the LoTSS ones derived from raw catalogues (see \citealt{ocran20} counts in top panel of Fig. \ref{fig:source_count_raw_debl}), indicating that the catalogue post-processing undertaken for the LoTSS deep fields  did play a role here. Indeed we could count on very complete identification rates ($\ga 97\%$ of radio sources have an optical/IR counterpart), enabling a very accurate process of association, deblending and artifact removal. This is not true for other samples which went through similar procedures, like the \citet{prandoni18} sample (identification rate $\sim 80\%$ down to S$_{\rm 1.4\, GHz}=0.12$ mJy) or the \citet{ocran20} sample (identification rate $\sim 90\%$). On the other hand, COSMOS counts may suffer from flux losses at their faintest end, being derived from a higher frequency (3 GHz), higher (sub-arcsec) resolution survey. Another source of systematics may be the  extrapolation of the higher frequency counts to 150 MHz, which is sensitive to the assumed spectral index. In particular, adopting the same spectral index value for all the sources may not be appropriate. There are indeed indications that the spectral index may flatten at 
mJy/sub-mJy regimes, due to the emergence of a population of low power, core dominated AGN (e.g. \citealt{prandoni06,Whittam13}) and re-steepen again at $\mu$Jy levels, where star-forming galaxies become dominant (e.g. \citealt{owen09}).

\vspace{0.5cm}
\subsection*{Qualitative comparison with models}


\normalfont
The source counts derived from the LoTSS Deep Fields provide unprecedented observational constraints to the shape of the  source counts at 150 MHz sub-mJy flux densities. As such they can be compared with counts predictions based on existing evolutionary models of radio source populations. A comprehensive comparison with models is beyond the scope of this paper, and will be the subject of forthcoming papers, where counts and luminosity functions will be presented and discussed for various radio source populations. Here we only provide some first qualitative considerations.

In Figs. \ref{fig:source_count_raw_debl} and \ref{fig:source_count_all} we compare the LoTSS source counts to the 150 MHz determinations derived from the  \citet{wilman08} and  \citet{bonaldi19} simulated catalogues\footnote{\citet{bonaldi19} present three simulated catalogues, each covering a different area of the sky. The one used here is the so-called {\it medium} tier, which  covers a 25 deg$^2$ field of view, providing a very good match to the LoTSS Deep Fields. 
We use a new version of the catalogues originally presented in \citet{bonaldi19}, which better reproduce the observations at the bottom and top of the covered frequency range (Bonaldi, private communication).} 
(black and dark violet shaded curves, respectively), as well as  from  \citet{mancuso17} models (light blue curve). We notice that the \citet{bonaldi19} and \citet{wilman08} source counts are very similar at the bright end and better reproduce the observations than \citet{mancuso17}. On the other hand, \citet{bonaldi19} and \citet{mancuso17} counts are very similar at the faint end, and in better agreement with the observations than \citet{wilman08}. Nevertheless, neither \citet{bonaldi19} nor \citet{mancuso17} can reproduce the pronounced bump at sub-mJy flux densities, observed in the counts derived from final catalogues (see Fig. \ref{fig:source_count_raw_debl}, bottom panel, or Fig. \ref{fig:source_count_all}). In addition all models appear to over-estimate the counts derived from our final catalogues at intermediate flux densities ($S\sim 2-20$ mJy). An over-prediction of the observed counts over T-RECS and \citet{wilman08} simulations in the flux density range $3-12$ mJy was also noticed by \cite{siewert20} in their analysis of the HETDEX field of LoTSS-DR1 (\citealt{shimwell17,shimwell19}).

In an attempt to better understand where the evolutionary models may fail, we compare the observed source redshift distribution (using redshifts from paper IV) with those of the \citet{bonaldi19} simulated catalogue. We restrict this comparison to the EN1 field, as it is the deepest and has the most complete optical coverage among the three LoTSS Deep Fields. 
We limit our analysis to the flux density range $0.25$ mJy $\leq S_{150} \leq 20$ mJy. At lower flux densities the effects of the visibility function and of incompleteness cannot be neglected anymore; at larger flux densities the available source statistics is very sparse. In the flux density regime under consideration, we have $14,916$ sources; $14,432$ of them (97\%) have measured redshifts  (91\%  photometric and  9\% spectroscopic, see paper IV). 3\% of the sources do not have  redshift estimates, and are not included in the analysis. We believe the effect of neglecting these sources does not alter the main results of our analysis. Photometric redshifts can be considered  robust up to $z=1.5$ for galaxies ($\sigma_{\rm NMAD}=1.48\times \rm median(|\Delta z| /(1+z_{\rm spec}))=0.02$), and up to  $z=4$ for AGN ($\sigma_{\rm NMAD}=0.064$; we refer to paper IV for more details).

\begin{figure*}[!thb]
\begin{center}
\resizebox{0.9\hsize}{!}{
\includegraphics[angle=0]{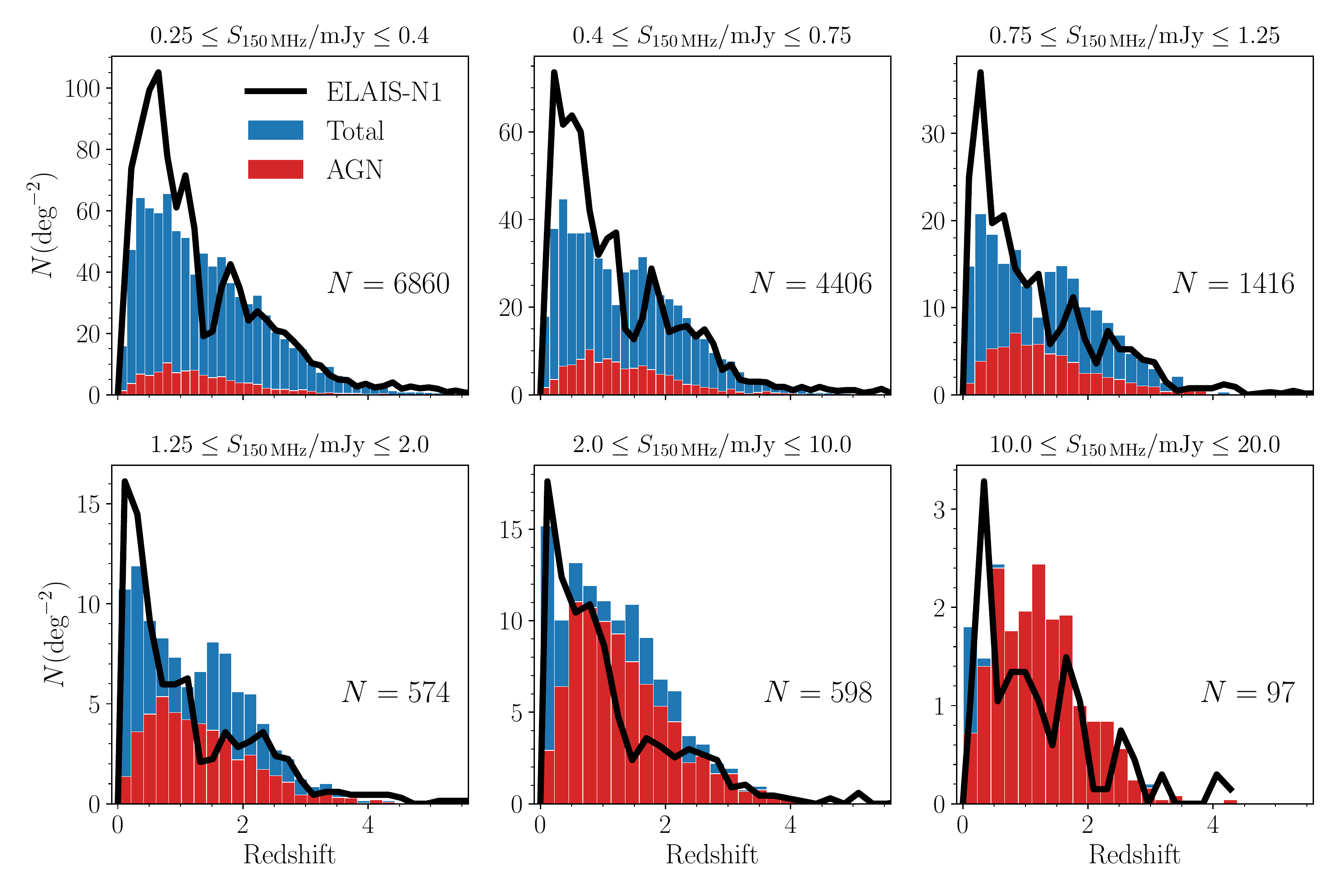}
}
\caption{Redshift distribution of the simulated sources in  \citet{bonaldi19} catalogue: the blue histogram corresponds to the total number of sources; the red histogram corresponds to the AGN component only. The solid black line shows the redshift distribution of the sources in the EN1 field. For a proper comparison the y-axis represents the source density in each catalogue. Each panel corresponds to a different flux density bin, increasing from left to right and from top to bottom. Redshift bins range from $\Delta z = 0.14$ to $\Delta z = 0.22$ from the faintest to the brightest flux bin, i.e. they are larger than the estimated photometric redshift scatter by a factor $2-3$ for AGN and by a factor $\geq 7$ for galaxies. We caution that the EN1 redshift distributions are reliable only up to $z\sim 1.5$ for galaxies and up to $z\sim 4$ for AGN.}
\label{fig:bonaldi-comparison}
\end{center}
\end{figure*}

In Fig. \ref{fig:bonaldi-comparison} we show the redshift distributions of the sources in the EN1 field (solid black lines) for various flux density bins. These distributions are compared with the simulated distributions based on  \citet{bonaldi19} evolutionary models (blue histogram bars). In the flux density bins spanning from 0.25 to 0.75 mJy we see a clear deficiency of the simulated sources, in agreement with the observed excess in the counts. This deficiency is mainly associated with sources at $z<1$. The fact that the source counts' excess is observed in all of the LoTSS fields (see e.g. Fig. \ref{fig:source_count_raw_debl}) seems to rule out the case this is due to cosmic variance. We cannot exclude however, that the clustering properties assumed in the simulation may play a role in producing the observed discrepancy\footnote{We note, however, that a similar discrepancy is observed when comparing EN1 to the {\it wide} tier of the \citet{bonaldi19} simulations, where clustering is not implemented.}.  At larger flux densities ($S\ga 1$ mJy), we see an opposite trend, i.e. there is an excess of simulated sources with respect to observations. This is again consistent with the deficiency observed in our counts in the range $2-20$ mJy. This excess is associated with intermediate redshift ($1<z<2$) sources, and appears to be mostly due to an excess of AGN in the simulated catalogue, at least at flux densities larger than a few mJy. We note that both the source excess at sub-mJy fluxes and the  AGN deficiency at the brightest fluxes are observed at redshifts where the involved populations (galaxies and AGN in the former, AGN only in the latter) have robust photometric redshift estimates.

\section{Conclusions}
In this paper, we have presented the source number counts derived from the LoTSS Deep Fields: the Lockman Hole (LH), the Bo\"otes (Boo) and the Elais-N1 (EN1). With central rms noise levels of 22, 33, 17 $\mu$Jy beam$^{-1}$ the LH, Boo and EN1 fields are the deepest obtained so far at 150 MHz, allowing us to get unprecedented observational constraints to the shape of the source counts at 150 MHz sub-mJy flux densities.
We compared the source counts derived from the LoTSS deep fields  with other existing source-counts determinations from low- and high- frequency radio surveys, and state-of-the-art evolutionary models. Our counts are in broad agreement with those from the literature, and show the well known upturn at $\la$ few mJy, which indicates the emergence of the star forming galaxy population. More interestingly, our counts show for the first time a very pronounced drop around $S\sim 2$ mJy, which results in a prominent `bump' at sub-mJy flux densities.  Such a `drop and bump' feature was not observed in previous counts' determinations (neither at 150 MHz nor at higher frequency). While sample variance can play a role in explaining the observed discrepancies, we believe this is mostly the result of a careful analysis aimed at deblending confused sources from the radio source catalogues (see paper III). Our counts cannot be fully reproduced by any of the existing evolutionary models. From a qualitative comparison with \citet{bonaldi19} simulated catalogues, we find that the  sub-mJy `bump' appears to be associated with an excess of low-redshift ($z<1$) galaxies and/or AGN, while the drop at mJy flux densities seems to be due to a deficiency of AGN at redshifts $1<z<2$. A more in-depth investigation of these preliminary results  will be the subject of forthcoming papers, based on more comprehensive comparisons with models.

\section{Acknowledgements}
We thank the anonymous referee for constructive suggestions that allowed us to improve our draft. This paper is based (in part) on data obtained with the International LOFAR Telescope (ILT) under project code LC3\_008. LOFAR (\citealt{vanhaarlem13}) is the Low Frequency Array designed and constructed by ASTRON. It has observing, data processing, and data storage facilities in several countries, that are owned by various parties (each with their own funding sources), and that are collectively operated by the ILT foundation under a joint scientific policy. The ILT resources have benefitted from the following recent major funding sources: CNRS-INSU, Observatoire de Paris and Université d’Orléans, France; BMBF, MIWF-NRW, MPG, Germany; Science Foundation Ireland (SFI), Department of Business, Enterprise and Innovation (DBEI), Ireland; NWO, The Netherlands; The Science and Technology Facilities Council, UK ; Istituto Nazionale di Astrofisica (INAF), Italy. This paper is based on the data obtained with the International LOFAR Telescope (ILT). The Leiden LOFAR team acknowledge support from the ERC Advanced Investigator programme NewClusters 321271 and the VIDI research programme with project number 639.042.729, which is financed by the Netherlands Organisation for Scientific Research (NWO). AD acknowledges support by the BMBF Verbundforschung under the grant 05A17STA. The Jülich LOFAR Long Term Archive and the German LOFAR network are both coordinated and operated by the J\"ulich Supercomputing Centre (JSC), and computing resources on the supercomputer JUWELS at JSC were provided by the Gauss Centre for supercomputing e.V. (grant CHTB00) through the John von Neumann Institute for Computing (NIC).
TMS and DJS acknowledge the Research Training Group 1620 `Models of Gravity', supported by Deutsche Forschungsgemeinschaft (DFG) and support of the German Federal Ministry for Science and Research BMBF-Verbundforschungsprojekt D-LOFAR IV (grant number 05A17PBA). PNB is grateful for support from the UK STFC via grant ST/R000972/1. MJH acknowledges support from the UK Science and Technology Facilities Council (ST/R000905/1). IP and MB acknowledge support from INAF under PRIN SKA/CTA "FORECaST" and PRIN MAIN STREAM “SAuROS” projects, as well as from the Ministero degli Affari Esteri e della Cooperazione Internazionale - Direzione Generale per la Promozione del Sistema Paese Progetto di Grande Rilevanza ZA18GR02. JS is grateful for support from the UK STFC via grant ST/R000972/1. MJJ acknowledges support from the UK Science and Technology Facilities Council [ST/N000919/1] and the Oxford Hintze Centre for Astrophysical Surveys which is funded through generous support from the Hintze Family Charitable Foundation. RK acknowledges support from the Science and Technology Facilities Council (STFC) through an STFC studentship via grant ST/R504737/1.	WLW acknowledges support from the ERC Advanced Investigator programme NewClusters 321271. WLW also acknowledges support from the CAS-NWO programme for radio astronomy with project number 629.001.024, which is financed by the Netherlands Organisation for Scientific Research (NWO). KLE acknowledges financial support from the Dutch Science Organization (NWO) through TOP grant 614.001.351. This research made use of APLpy, an open-source plotting package for Python hosted at \href{url}{http://aplpy.github.com.}

\bibliographystyle{aa}
\bibliography{paper.bib} 

\clearpage
\onecolumn

\begin{appendix}

\section{LoTSS Deep Fields - Counts' tables}\label{sec:counts-tables}

\begin{table}[tbh]
    \caption{150~MHz normalized differential radio-source counts as derived from the Lockman Hole final catalogues. }
   \label{tab:counts_LH}
    \centering
   \begin{tabular}{|c|c|c|c|c|c|c|c|c|c|}
   \hline
      $S_{\rm{min}}$ & $S_{\rm{max}}$ & $\Delta S$ & x & $N_S$ & $N^{+\sigma}_{-\sigma}$ & $\rm{Sys}^-$ & $\rm{Sys}^+$ & $c_1$ & $c_2$    \\
         \hline
 0.25 & 0.35 & 0.10 & 0.30 & 6673 & $42.35_{0.52}^{0.52}$ & 4.56 & 2.82 & 1.12 & 1.34 \\
0.35 & 0.50 & 0.15 & 0.42 & 5314 & $ 48.08_{0.66}^{0.67}$ & 1.38 & 2.73 & 1.01 & 1.10  \\
0.50 & 0.71 & 0.21 & 0.59 & 3545 & $ 51.21_{0.86}^{0.87}$ & 0.24 & 2.48 & 1.00 & 1.06 \\
0.71 & 1.00 & 0.29 & 0.84 & 2077 & $ 49.18_{1.08}^{1.10}$ & 0.02 & 2.09 & 1.00 & 1.04 \\
1.00 & 1.41 & 0.41 & 1.19 & 1119 & $ 43.55_{1.30}^{1.34}$ & 0.04 & 1.54 & 1.01 & 1.03 \\
1.41 & 2.00 & 0.59 & 1.68 & 683 & $ 44.29_{1.70}^{1.76}$ & 0.00 & 1.89 & 1.01 & 1.02 \\
2.00 & 2.83 & 0.83 & 2.38 & 390 & $ 42.01_{2.13}^{2.24}$ & 0.00 & 1.94 & 1.01 & 1.01 \\
2.83 & 4.00 & 1.17 & 3.36 & 272 & $ 49.19_{2.98}^{3.16}$ & 0.00 & 2.77 & 1.01 & 1.01 \\
4.00 & 5.66 & 1.66 & 4.76 & 184 & $ 55.86_{4.12}^{4.42}$ & 0.00 & 3.60 & 1.01 & 1.01 \\
5.66 & 8.00 & 2.34 & 6.73 & 148 & $ 75.45_{6.20}^{6.71}$ & 0.00 & 4.09 & 1.01 & 1.01 \\
8.00 & 11.3 & 3.30 & 9.51 & 132 & $ 113.1_{9.9}^{10.7}$ & 0.0 & 4.0 & 1.01 & 1.01 \\
11.3 & 22.6 & 11.3 & 16.0 & 172 & $ 158.4_{12.1}^{13.0}$ & 0.0 & 2.3 & 1.01 & 1.01 \\
22.6 & 45.3 & 22.7 & 32.0 & 128 & $ 333.2_{29.5}^{32.1}$ & 0.0 & 1.3 & 1.01 & 1.01 \\
45.3 & 90.5 & 45.2 & 64.0 & 74 & $ 543.7_{63.2}^{70.6}$ & 0.2 & 2.2 & 1.01 & 1.01 \\
90.5 & 181 & 90.5 & 128 & 45 & $ 935.1_{139.4}^{160.2}$ & 2.5 & 4.2 & 1.01 & 1.01 \\
181 & 724 & 543 & 362 & 46 & $ 2144_{316}^{363}$ & 3 & 15 & 1.01 &  1.01 \\
724 & 2896 & 2172 & 1448 & 17 & $ 6335_{1537}^{1909}$ & 18 & 86 & 1.00 & 1.00 \\

      \hline
    \end{tabular}
    \tablefoot{Columns are as follows: $S_{\rm{min}}$ and $S_{\rm{max}}$ are the minimum and maximum flux densities (expressed in mJy), respectively; $\Delta$ denotes the flux density intervals; $x$ is the geometric mean of $S_{\rm{min}}$ and $S_{\rm{max}}$; $N_S$ is the number of sources in respective bins; $N$ is the normalized source counts and $\sigma$ is the Poissonian errors on the normalized counts; $\rm{Sys}^+$ and $\rm{Sys}^-$ are the systematic errors, accounting for different modeling of resolution and Eddington bias corrections (see Sects. \ref{sec:corrresbias} and \ref{sec:eddbias} for more details); the correction factor ${c_1}$ represents the weighting applied to the counts to account for resolution and Eddington biases; the correction factor $c_2$ also includes the weighting due to the visibility function.}
\end{table}

\begin{table}[tbh]
    \caption{150~MHz normalized differential radio-source counts as derived from the Bo\"otes final catalogues. Parameters as in Tab. \ref{tab:counts_LH}}.
   \label{tab:counts_Boo}
    \centering
   \begin{tabular}{|c|c|c|c|c|c|c|c|c|c|}
   \hline
      $S_{\rm{min}}$ & $S_{\rm{max}}$ & $\Delta S$ & x & $N_S$ & $N^{+\sigma}_{-\sigma}$ & $\rm{Sys}^-$ & $\rm{Sys}^+$ & $c_1$ & $c_2$\\
         \hline
0.35 & 0.50 & 0.15 & 0.42 & 3992 & $ 46.96_{0.74}^{0.76}$ & 1.91 & 3.38 & 1.04 & 1.19 \\
0.50 & 0.70 & 0.20 & 0.59 & 3066 & $ 53.73_{0.97}^{0.99}$ & 0.23 & 3.39 & 1.02 & 1.08 \\
0.70 & 1.00 & 0.30 & 0.84 & 1812 & $ 51.33_{1.21}^{1.23}$ & 0.00 & 2.76 & 1.02 & 1.06 \\
1.00 & 1.41 & 0.41 & 1.19 & 949 & $ 44.17_{1.43}^{1.48}$ & 0.00 & 2.01 & 1.02 & 1.03 \\
1.41 & 1.99 & 0.58 & 1.68 & 580 & $ 44.64_{1.85}^{1.93}$ & 0.00 & 2.55 & 1.02 & 1.02 \\
1.99 & 2.82 & 0.83 & 2.37 & 345 & $ 44.35_{2.39}^{2.52}$ & 0.00 & 2.94 & 1.02 & 1.02 \\
2.82 & 3.99 & 1.17 & 3.35 & 234 & $ 50.83_{3.32}^{3.54}$ & 0.00 & 3.98 & 1.01 & 1.02 \\
3.99 & 5.64 & 1.65 & 4.74 & 155 & $ 56.20_{4.51}^{4.88}$ & 0.00 & 4.99 & 1.01 & 1.02 \\
5.64 & 7.97 & 2.33 & 6.70 & 146 & $ 88.87_{7.36}^{7.96}$ & 0.00 & 6.99 & 1.01 & 1.01 \\
7.97 & 11.3 & 3.33 & 9.48 & 117 & $ 119.8_{11.1}^{12.1}$ & 0.0 & 5.9 & 1.01 & 1.01 \\
11.3 & 22.6 & 11.3 & 15.9 & 125 & $ 137.4_{12.3}^{13.4}$ & 0.0 & 2.8 & 1.01 & 1.01 \\
22.6 & 45.1 & 22.5 & 31.9 & 96 & $ 298.1_{30.4}^{33.5}$ & 0.0 & 1.4 & 1.01 & 1.01 \\
45.1 & 90.2 & 45.1 & 63.8 & 75 & $ 657.5_{75.9}^{84.7}$ & 0.8 & 3.2 & 1.01 & 1.01 \\
90.2 & 180 & 89.8 & 128 & 41 & $ 1018_{159}^{184}$ & 6 & 6 & 1.01 & 1.01 \\
180 & 722 & 542 & 361 & 27 & $ 1502_{289}^{345}$ & 3 & 12 & 1.01 & 1.01 \\
722 & 2886 & 2164 & 1443 & 11 & $ 4861_{1466}^{1908}$ & 9 & 48 & 1.00 & 1.00 \\
      \hline
    \end{tabular}
\end{table}

\begin{table}[tbh]
    \caption{150~MHz normalized differential radio-source counts as derived from the ELAIS-N1 final catalogues. Parameters as in Tab. \ref{tab:counts_LH}}.
   \label{tab:counts_EN1}
    \centering
   \begin{tabular}{|c|c|c|c|c|c|c|c|c|c|}
   \hline
      $S_{\rm{min}}$ & $S_{\rm{max}}$ & $\Delta S$ & x & $N_S$ & $N^{+\sigma}_{-\sigma}$ & $\rm{Sys}^-$ & $\rm{Sys}^+$ & $c_1$ & $c_2$\\
         \hline
0.18 & 0.26 & 0.08 & 0.22 & 7187 & $ 38.85_{0.46}^{0.46}$ & 6.69 & 2.52 & 1.16 & 1.29 \\
0.26 & 0.36 & 0.10 & 0.31 & 5340 & $ 43.54_{0.60}^{0.60}$ & 2.11 & 2.33 & 1.02 & 1.08 \\
0.36 & 0.51 & 0.15 & 0.43 & 3397 & $ 44.84_{0.77}^{0.78}$ & 0.63 & 1.97 & 1.00 & 1.03 \\
0.51 & 0.73 & 0.22 & 0.61 & 2081 & $ 45.40_{1.00}^{1.02}$ & 0.02 & 1.72 & 1.00 & 1.01 \\
0.73 & 1.03 & 0.30 & 0.86 & 1165 & $ 42.58_{1.25}^{1.28}$ & 0.00 & 1.32 & 1.00 & 1.01 \\
1.03 & 1.45 & 0.42 & 1.22 & 676 & $ 41.18_{1.58}^{1.65}$ & 0.00 & 1.07 & 1.01 & 1.01 \\
1.45 & 2.06 & 0.61 & 1.73 & 368 & $ 37.59_{1.96}^{2.06}$ & 0.00 & 1.01 & 1.01 & 1.01 \\
2.06 & 2.91 & 0.85 & 2.45 & 232 & $ 39.83_{2.62}^{2.79}$ & 0.00 & 1.35 & 1.01 & 1.01 \\
2.91 & 4.11 & 1.20 & 3.46 & 163 & $ 47.00_{3.68}^{3.97}$ & 0.00 & 2.05 & 1.01 & 1.01 \\
4.11 & 5.82 & 1.71 & 4.89 & 100 & $ 48.44_{4.84}^{5.33}$ & 0.00 & 2.41 & 1.01 & 1.01 \\
5.82 & 8.23 & 2.41 & 6.92 & 86 & $ 69.90_{7.54}^{8.35}$ & 0.00 & 2.46 & 1.00 & 1.00 \\
8.23 & 11.6 & 3.37 & 9.79 & 66 & $ 90.85_{11.18}^{12.56}$ & 0.00 & 2.35 & 1.01 & 1.01 \\
11.6 & 23.3 & 11.7 & 16.5 & 89 & $ 130.7_{13.9}^{15.3}$ & 0.0 & 1.0 & 1.00 & 1.00 \\
23.3 & 46.6 & 23.3 & 32.9 & 93 & $ 385.9_{40.0}^{44.2}$ & 0.0 & 1.2 & 1.00 & 1.00 \\
46.6 & 93.1 & 46.5 & 65.8 & 54 & $ 634.4_{86.3}^{98.1}$ & 0.4 & 2.1 & 1.00 & 1.00 \\
93.1 & 186 & 92.9 & 132 & 37 & $ 1230_{202}^{235}$ & 3 & 5 & 1.00 & 1.00 \\
186 & 745 & 559 & 372 & 17 & $ 1271_{308}^{383}$ & 2 & 8 & 1.01 & 1.01 \\
745 & 2979 & 2234 & 1489 & 5 & $ 2980_{1333}^{1929}$ & 7 & 34 & 1.00 & 1.00 \\
      \hline
    \end{tabular}
\end{table}

\end{appendix}

\label{lastpage}
\end{document}